\documentclass[range]{ar2e}
\usepackage{ulem}  % For Copy Editors
\usepackage{amsmath}
\usepackage{amsfonts}
\usepackage{amssymb}

\begin{document}
\input epsf.tex    %<-If you need EPS figures to be
                   %  called in {figure} environment for PC
\input psfig.sty

\renewcommand\refname{LITERATURE CITED}

\font\tenimbf=cmmib10 at 10pt
\font\sevenimbf=cmmib10 at 6pt
\font\fiveimbf=cmmib10 at 4pt
%\font\tenimbf=cmmib10 at 12pt
%\font\sevenimbf=cmmib10 at 7pt
%\font\fiveimbf=cmmib10 at 5pt
\newfam\imbf
\textfont\imbf=\tenimbf
\scriptfont\imbf=\sevenimbf
\scriptscriptfont\imbf=\fiveimbf
\def\imb{\fam\imbf\tenimbf}

\newcommand{\be}{\begin{equation}}
\newcommand{\bea}{\begin{eqnarray}}
\newcommand{\ee}{\end{equation}}
\newcommand{\eea}{\end{eqnarray}}

%%%%%%%%%%%%%%%%%%%%%%%_eraita_vektoreita_%%%%%%%%%%%%%%%%%%%%%
%%%%%%%%

\newcommand{\ba}{{\mathbf a}}              %% These are roman bold
\newcommand{\bbb}{{\mathbf b}}
\newcommand{\bc}{{\mathbf c}}
\newcommand{\bbe}{{\mathbf e}}
\newcommand{\bbf}{{\mathbf f}}
\newcommand{\bi}{{\mathbf i}}
\newcommand{\bj}{{\mathbf j}}
\newcommand{\bk}{{\mathbf k}}
\newcommand{\bn}{{\mathbf n}}
\newcommand{\bp}{{\mathbf p}}
\newcommand{\br}{{\mathbf r}}
\newcommand{\hbr}{{\mathbf \hat r}}
\newcommand{\bs}{{\mathbf s}}
\newcommand{\bu}{{\mathbf u}}
\newcommand{\bv}{{\mathbf v}}
\newcommand{\bw}{{\mathbf w}}
\newcommand{\bx}{{\mathbf x}}
\newcommand{\by}{{\mathbf y}}
%%%%%%%%%%%%%%%%%%%%%%_uppercase_%%%%%%%%%%%%%%%%%%%%%%%%%%%%%%
%%%%%%
\newcommand{\bA}{{\mathbf $A$}}
\newcommand{\bB}{{\mathbf $B$}}
\newcommand{\bC}{{\mathbf $C$}}
\newcommand{\bD}{{\mathbf $D$}}
\newcommand{\hbD}{{\mathbf $\hat D$}}
\newcommand{\bE}{{\mathbf $E$}}
\newcommand{\bF}{{\mathbf $F$}}
\newcommand{\bP}{{\mathbf $P$}}
\newcommand{\bR}{{\mathbf $R$}}
\newcommand{\hbR}{{\mathbf $\hat R$}}

\newcommand{\ib}{\mbox{\boldmath $i$}}   %%these are math italic
\newcommand{\jb}{\mbox{\boldmath $j$}}   %%bold
\newcommand{\kb}{\mbox{\boldmath $k$}}
\newcommand{\xb}{\mbox{\boldmath $x$}}
\newcommand{\yb}{\mbox{\boldmath $y$}}
\newcommand{\ub}{\mbox{\boldmath $u$}}
\newcommand{\vb}{\mbox{\boldmath $v$}}
\newcommand{\wb}{\mbox{\boldmath $w$}}
\newcommand{\rb}{\mbox{\boldmath $r$}}
\newcommand{\hrb}{\mbox{\boldmath $\hat r$}}
\newcommand{\ab}{\mbox{\boldmath $a$}}
\newcommand{\bb}{\mbox{\boldmath $b$}}
\newcommand{\cb}{\mbox{\boldmath $c$}}
\newcommand{\eb}{\mbox{\boldmath $e$}}
\newcommand{\nb}{\mbox{\boldmath $n$}}
\newcommand{\sbb}{\mbox{\boldmath $s$}}
\newcommand{\fb}{\mbox{\boldmath $f$}}
\newcommand{\pb}{\mbox{\boldmath $p$}}
%%%%%%%%%%%%%%%%%%%%%%_uppercase_%%%%%%%%%%%%%%%%%%%%%%%%%%%%%%
%%%%%%
\newcommand{\Fb}{\mbox{\boldmath $F$}}
\newcommand{\Pb}{\mbox{\boldmath $P$}}
\newcommand{\Rb}{\mbox{\boldmath $R$}}
\newcommand{\hRb}{\mbox{\boldmath $\hat R$}}
\newcommand{\Eb}{\mbox{\boldmath $E$}}
\newcommand{\Bb}{\mbox{\boldmath $B$}}
\newcommand{\Ab}{\mbox{\boldmath $A$}}
\newcommand{\Cb}{\mbox{\boldmath $C$}}
\newcommand{\Db}{\mbox{\boldmath $D$}}
\newcommand{\hDb}{\mbox{\boldmath $\hat D$}}

\def\Teff{T_{\rm eff}}
\def\psat{p_{\rm sat}}
\def\avpt{\langle p_{\rm T}\rangle}
\def\ssNN{\sqrt{s_{NN}}}
\def\av#1{\langle#1\rangle}
\def\half{$\scriptstyle{1\over 2}$}
\def\mhalf{\scriptstyle{1\over 2}}
\def\Teff{T_{\rm eff}}
\def\tdec{\tau_{\rm dec}}
\def\Tdec{T_{\rm dec}}
\def\eosa{EoS$\,$A}
\def\eosp{EoS$\,$P}
\newcommand{\lsim}
{{\;\raise0.3ex\hbox{$<$\kern-0.75em\raise-1.1ex\hbox{$\sim$}}\;}}
\newcommand{\gsim}
{{\;\raise0.3ex\hbox{$>$\kern-0.75em\raise-1.1ex\hbox{$\sim$}}\;}}

\jname{Annu. Rev. Nucl. Particle Science}
\jyear{2006}
\jvol{56}
\ARinfo{1056-8700/97/0610-00}

\title{Hydrodynamic Models for Heavy Ion Collisions\thanks{Final 
       version of this review is scheduled to appear in the Annual
       Review of Nuclear and Particle Science Vol.\ 56, to be published
       in November 2006 by Annual Reviews, see ``www.annualreviews.org''.}}

\markboth{P.~Huovinen and P.V.~Ruuskanen}{Hydrodynamic Models
for Heavy Ion Collisions}

\author{P. Huovinen
\affiliation{Department of Physics, University of Virginia,
Charlottesville, VA 22904, USA\\
\textit{and}\\
Helsinki Institute of Physics, FIN-00014 University of Helsinki, Finland\\
{\it email: ph4h@virginia.edu}}
P.V. Ruuskanen
\affiliation{Department of Physics, FIN-40014 University of
Jyv\"askyl\"a, Finland\\
{\it email: vesa.ruuskanen@phys.jyu.fi}}}


\begin{abstract}
Application of hydrodynamics for modeling of heavy-ion collisions is
reviewed. We consider several physical observables that can be
calculated in this approach and compare them to the experimental
measurements.
\end{abstract}

\maketitle
\section{INTRODUCTION}
  \label{intro}

Quantum chromodynamics (QCD), the theory on strong interactions, has
been tested extensively in hard, large-momentum processes. In these
processes, a large amount of energy or three-momentum is transferred
to one (e.g., deep inelastic scattering) or few (jet production) quark
and gluon constituents of initial hadrons. Owing to the asymptotic
freedom property of QCD, the coupling strength is small.  The time
scale for this part is a small fraction of the time scale of the
overall process. These two properties allow the factorization and
perturbative treatment of the hard part of the process from the rest
of the matrix element. This programme has been very successful, and no
clear discrepancies have been found between experimental results and
theoretical calculations.

The goal of the experimental heavy ion program at ultra-relativistic
energies is to study QCD in an environment very different from that
encountered in hard processes, in a dense system of quarks and gluons.
When the heavy ion programme started two decades ago, the original
goals were the production and study of dense, thermally equilibrated,
strongly interacting matter, the quark-gluon plasma (QGP). Although
this is still the highest priority, phenomena other than the formation
of thermally equilibrated QGP also occur in dense partonic systems and
can be studied in heavy ion collisions. E.g.\ the existence of what is
called a Color Glass Condensate (CGC), which is related to the
saturation of gluon occupation numbers in dense components of initial
wave functions, may be important for the formation of the QGP and may
also have other observable effects~\cite{McLerran:1999hj}. Also, the
phenomenon of jet quenching, the loss of energy of a high-energy
parton (quark or gluon) when it traverses a high-density parton
system, can be expected to occur in the dense environment of heavy ion
collisions, even when the system is not fully equilibrated.

The earlier experiments at Brookhaven National Laboratory (BNL) and
CERN-SPS provided clear evidence of collective phenomena in nuclear
collisions. Recent results from experiments at the Relativistic Heavy
Ion Collider (RHIC) at BNL show that total multiplicities exceeding
1000 particles per unit rapidity, $dN/dy\gsim 1000$, are produced in
central (head-on) collisions of gold nuclei at center-of-mass energies
up to $\sqrt{s_{NN}} = 200$ GeV. Here $\sqrt{s_{NN}}$ is the
center-of-mass energy for a nucleon pair. The measured average
transverse momentum is $ p_T\gsim 0.5$ GeV, indicating a total energy
per unit rapidity interval of $dE_T/dy\gsim 500$ GeV. At time $\tau$
after the nuclei have passed through one another, the volume occupied
by the produced quanta in a rapidity interval $\Delta y$ is $\Delta
V~=~\tau\Delta yA_T$, or $\tau A_T=\tau\pi R_A^2$ for unit
rapidity. At $\tau=1$ fm/$c$ the above numbers
imply~\cite{Bjorken:1982qr} a particle density of $\sim 10$~fm$^{-3}$
and an energy density of $\sim 5$~GeVfm$^{-3}$, well above different
estimates and the results from lattice calculations of the energy
density at which the phase transition from confined hadrons to
unconfined quarks and gluons occur.

The subject of this review is the use of hydrodynamic modeling to
describe the expansion and dilution of matter produced in nuclear
collisions. One motivation for this is to formulate a framework to
study different observable quantities and correlations among them.
Obviously hydrodynamics alone does not suffice because, first, at high
energies, particle production can not be included and, second, only
the properties of produced particles, not the hydrodynamic densities
during the expansion, can be measured directly. Below we describe a
possible dynamic scenario to calculate the production of initial
matter. This provides the initial conditions for solving the
hydrodynamic equations. Because the production dynamics is still not
completely under control, sometimes it may be useful to use physically
motivated parametrizations of initial conditions to study, for
example, how the details of particle spectra depend on different
features of initial conditions. The hydrodynamic description also
needs another supplement. A link from hydrodynamical quantities to
particle spectra is necessary at the end of expansion, when the
particles become independent and fly to detectors.

The use of hydrodynamic concepts like temperature, pressure, and flow
velocity cannot be strictly justified for matter formed in a heavy ion
collision. Although the total number of particles produced is several
thousand at RHIC and may be an order of magnitude more in the future
ALICE experiments at the Large Hadron Collider (LHC) at CERN, they
hardly form a macroscopic system in the proper thermodynamic
sense. However, one can argue for partial equilibration and the
formation of collective phenomena from the numbers given above. At
initial particle densities of the order of 5--10~fm$^{-3}$, and even
with modest estimates of cross sections of the order of 1--2~mb, mean
free paths are $\lsim 1$ fm, much below the nuclear size $2R_A \gsim 10$
fm. Therefore, frequent collisions occur and momentum is transferred
from denser regions toward less dense regions. In describing the main
consequences of these secondary collisions, concepts like temperature
and pressure are useful.

Although the use of hydrodynamics may be justified for some features
of nuclear collisions, that need not be the case for others.  Defining
energy density and pressure in the usual way in terms of the local
momentum distribution, one obtains for pressure of massless particles
$P=\epsilon/3$ for any isotropic momentum distribution $f(p),\ p=|\pb|=E$.
In this case, describing the build-up of collective motion or flow
using hydrodynamics with $P=\epsilon/3$ as the equation of state (EoS)
could be a reasonable approach, even when the form of $p$ dependence
of $f(p)$ differs from that of the thermal equilibrium distribution,
$f_{\rm th}(p)$. However, if $f(p)$ differs significantly from
$f_{\rm th}(p)$, any conclusions based on the detailed momentum
dependence of $f(p)$ would fail. This could well be the case at large
momenta. Initial production is expected to contain a component that
has an approximate power-law behavior at large momenta. It will take
longer and requires a larger volume than available to change this
power behavior to the exponential form of thermal distribution. The
observed behavior of high-$p_T$ hadron spectra shows clearly that the
high-momentum partons are not thermalized and that they suffer an
energy loss while traversing the produced dense matter. Because the
high-energy partons form only a small part of produced matter both in
multiplicity and in transverse energy, a thermal equilibrium
description can still be adequate for the bulk of the matter.

Keeping in mind all the reservations, we review the use of
hydrodynamics in describing the heavy ion collisions at collider
energies, and the calculation of observable quantities. We compare the
calculated results to the measurements mainly at RHIC, mentioning some
results from the CERN-SPS. We also present examples of predictions for
the future ALICE experiment at CERN.

%%%%%%%%%%%%%%%%%%%%%%%%%%%%%%%%%%%%%%%%%%%%%%%%%%%%%%%%%%%%%%%
%%%%%%%%%%%
\section{HYDRODYNAMIC EXPANSION}
\label{sec:hydro}
\def\Tmn{T^{\mu\nu}}
%%%%%%%%%%%%%%%%%%%%%%%%%%%%%%%%%%%%%%%%%%%%%%%%%%%%%%%%%%%%%%%
%%%%%%%%%%%

Hydrodynamics is the theoretical framework describing the motion of
fluid, a continuous, flowing medium. The equation of motion can be
derived from kinetic equations. Hydrodynamic equations take the
simplest form if local thermal equilibrium is assumed. In this
treatment, there are no dissipative effects. To take such effects into
account by approximation, small deviations from the local equilibrium
are assumed. A linear treatment of deviations leads to a system of
equations which contains the viscous coefficients including heat
conductivity, in the case of conserved currents. For a review, see
Reference \cite{Rischke}. In most applications of hydrodynamics to
heavy ion collisions, viscosity has been neglected. In studies with
viscosity, results on global, integrated quantities do not differ
qualitatively from those without viscosity. However, for example,
transverse spectra at larger transverse momenta $p_T\gsim 1.5$ GeV may
start to deviate clearly from ideal-fluid
calculation~\cite{Teaney-visco,Muronga,Muroya:2004pu}. One should
remember, though, that the viscous properties of strongly interacting
matter are not well understood, and the approximations in the
numerical work also introduce uncertainties.

Once EoS is known and initial conditions are specified, the
hydrodynamic equations determine the expansion of the fluid. In the
context of describing heavy ion collisions, the use of these equations
requires knowing the EoS of strongly interacting matter and the
primary production of particles. Detailed knowledge of microscopic
processes is not required if a very strong assumption is taken: The
expanding system stays in local thermodynamical equilibrium. This
becomes of great practical importance if one wants to include in the
hydrodynamic expansion the transition from quarks and gluons to
hadrons. The complicated deconfinement or hadronization processes need
not be known in microscopic detail; all that is necessary is the
thermodynamic equation of state as computed, for example, in lattice
QCD.

\subsection{Hydrodynamic Equations}
The hydrodynamic equations
\be
\partial_\mu T^{\mu\nu}=0\,,\quad
T^{\mu\nu}=(\epsilon+P)u^\mu u^\nu - P\,g^{\mu\nu}
\label{eq:hydro}
\ee
express, in terms of the energy-momentum tensor $T^{\mu\nu}$, the
conservation of energy and momentum in continuous, flowing matter.
The quantities defining $\Tmn$ are: $\epsilon:=$ the energy density,
$P:=$ the pressure, and $u^\mu:=$ the flow four-velocity, normalized
to $u_\mu u^\mu=1$ as usual. The simple form of $\Tmn$ above holds for
an ideal fluid.

In addition, if the system contains conserved densities $n_i$, such as
those of charge and baryon number, their evolution is expressed by
continuity equations
\be
\partial_\mu j_i^\mu = 0\,.
\label{eq:conscur}
\ee
With several conserved densities or non-zero viscous terms, the
definition of $u^\mu$ is not unique. One can define the velocity in
terms of one of the conserved currents by writing $j_i^\mu = n_iu_i^\mu$,
where $n_i=\sqrt{j_{i\mu}j_i^\mu}$, or one can use the energy-momentum
tensor to define the flow velocity. In the first case, usually
referred to as the Eckart definition, there is no flow of charge $Q_i$
in the local rest frame $u_i^\mu=(1,0)$, the Eckart frame. For a
non-ideal fluid, the energy flow would usually be non-zero in the
Eckart frame. A definition of the fluid velocity in terms of the
energy-momentum tensor, referred to as the Landau definition, is such
that the energy flow is zero in the local rest frame, but usually the
flow of different charges does not vanish.  Here, the only charge we
consider is the baryon density. We also treat the matter in the final
state as an ideal fluid, and thus the two choices for the flow
velocity coincide. For a review, see Reference~\cite{Rischke}.

The only properties of the dynamics contained in Equations
\ref{eq:hydro} and \ref{eq:conscur} are the conservation laws.
However, the relations between the thermodynamic densities $\epsilon$,
$P$ and $n_i$, or alternatively their definitions in terms of
temperature $T$ and chemical potentials $\mu_i$ --- e.g.,
$\epsilon=\epsilon(T,\mu_i)$ --- constituting the EoS, depend on the
details of the dynamics among the constituents of the matter. The need
for an EoS is obvious: The Equations \ref{eq:hydro} and
\ref{eq:conscur} contain five equations whereas there are six
quantities to be defined by solving the equations. These quantities
are the three components of the velocity, the energy density, pressure
and the baryon-number density. For a non-ideal flow, transport
coefficients would enter into the expressions of the energy-momentum
tensor and currents, and their derivation from theory requires the
knowledge of microscopic dynamics in the same way as the derivation of
EoS.

Most of the detailed hydrodynamical discussion below is limited to the
situation of scaling longitudinal flow and invariance under
longitudinal Lorentz boosts. This means that the longitudinal flow
velocity is $v_z=z/t$, and hence, the flow rapidity is
$\eta=\log[(t+z)/(t-z)]$, which is also often termed the space-time
rapidity. If the initial densities are assumed to depend on $t$ and
$z$ only through the longitudinal proper time $\tau=\sqrt{t^2-z^2}$,
e.g., $\epsilon=\epsilon(\tau,r)$, the expansion will evolve so that
densities remain independent of $\eta$ and the $v_z$ will retain the
scaling form $v_z=z/t$. In this situation the system is said to be
boost-invariant~\cite{Bjorken:1982qr}.

Before discussing the EoS in more detail, we note that the most useful
form of the EoS for solving the hydrodynamical equations is provided
by the relations among the densities $\epsilon$, $P$, and $n_i$ when
the hydrodynamical equations are written in the form of
Equations~\ref{eq:hydro} and \ref{eq:conscur}. In this form,
temperature and chemical potentials do not appear in the
equations. For the calculation of observable quantities, such as
hadron spectra or electromagnetic emission, relations among the above
densities and the temperature and chemical potentials must be
specified, as we discuss below.

The other ingredient that must be provided from outside into the
hydrodynamic description are the initial conditions, e.g., in terms of
initial energy distribution and velocity. From the physics point of
view this is a very profound problem because it involves not only the
primary production dynamics, but also the question of thermalization
of produced particles. We discuss first the initial conditions, then
the EoS, and finally the calculation of physical observables.

\subsection{Initial Conditions}
\label{sec:inicond}

As mentioned above, primary particle production cannot be formulated
within the hydrodynamic framework in a realistic way in high-energy
nuclear collisions. The dynamics of particle production is a separate
problem and, if solved, it provides the initial conditions for the
hydrodynamic expansion. Initial conditions specify the thermodynamic
state of the matter and its velocity on an appropriate space-time
boundary, which, for example, in the boost-invariant case discussed
below, can be taken to be a constant, longitudinal proper time surface
$\tau=\sqrt{t^2-z^2}=\tau_0$. From the point of view of the
hydrodynamical calculation, the initial conditions can be provided
either by a dynamical calculation of primary particle production or by
a reasonable parametrization, with the parameters either given by
physical arguments or fixed by comparing (some of) the results with
experimental data.

There are different approaches to primary production, such as pQCD +
final-state saturation (minijet) \cite{Eskola:2001dd} and the color
glass condensate model \cite{McLerran:1993ni} based on the initial
state parton saturation \cite{Gribov:tu}. Both describe the produced
matter as a parton system. Also, models based on string formation and
decay, such as the DPMJET model \cite{Ranft:fd}, are used for the
calculation of final hadron spectra. In such a model, a varying
fraction of energy is in the form of color strings at an early stage
of the collision, and they are not readily connected with a
hydrodynamic description that assumes particle-like constituents of
matter. We do not consider string models further, but we describe
briefly parton-based approaches to the primary production.

%%%%%%%%%%%%

To illustrate different key factors that enter the determination of
initial conditions from dynamical calculation of particle production,
we consider a perturbative QCD calculation of parton production as an
example. For such a calculation to converge, a cut-off must be
provided on small momentum-transfer collisions. In this model the
cut-off is obtained from a saturation condition expressed in terms of
the transverse nuclear geometry and the number of produced partons.
At collider energies, the saturation scale turns out to be typically
$\psat~\sim~1\ldots3$ GeV, which is clearly larger than
$\Lambda_{\rm QCD}\sim 0.2$ GeV. Because this cut-off is smaller than
what is usually used in jet calculations, produced partons are often
called minijets, as partons close to the cut-off dominate the
production. In addition to the cut-off, the ingredients of the
calculation are the parton distribution functions of colliding nuclei
and the parton-parton cross sections. These cross sections can be
calculated from basic QCD theory, but parton distributions must be
provided from other measurements.

The nuclear parton distributions are usually expressed in terms of
parton distributions of nucleons that, however, are known to be
modified in nuclei. The nuclear modification factor $R_A(x,Q^2)$ is
the nuclear parton distribution normalized to a single nucleon and
divided by the parton distribution of a free nucleon.\footnote{In the
actual calculation protons and neutrons are treated separately.}

The perturbative QCD calculation of minijet production is a
momentum-space calculation, as is the case in most production models.
To define the initial spatial densities, a connection between the
momentum of a produced parton and its space-time formation point is
needed. At collider energies, the hard partons of the colliding nuclei
are Lorentz contracted to a region on order of $2R_A/\gamma_{\rm cm}<< 1$
fm. We consider the collision region as a point in the longitudinal
direction that allows us to assume that the rapidity of the minijet
coincides with the space--time rapidity of the formation point,
$y=\eta=(1/2)\ln[(t+z)/(t-z)]$. We take the formation (proper) time to
be the inverse of the saturation scale, $\tau_0=1/p_{\rm sat}$.  Thus,
the minijet matter forms along the hyperbola $t=\sqrt{z^2+\tau_0^2}$
with initial longitudinal flow velocity $v_z(\tau_0)=z/t$. To
determine the transverse distribution, we must start with a
calculation of production cross section.

To obtain the initial conditions for baryon-number density and energy
density, we first need the minijet cross sections for (anti)quarks and
gluons and their first moments in transverse energy (momentum) in
nucleon-nucleon collision, each calculated in a rapidity interval
$\Delta y$ and integrated in $p_T$ from the saturation cut-off
$p_T=\psat$ to its maximum value
\cite{Eskola:2000fc,Eskola:1988yh}:
  \[
  \begin{split}
\sigma_{\rm jet}(\psat,\sqrt s,\Delta y,A)
&=\int_{\psat}^{\sqrt s/2} dp_T
\frac{d\sigma_{\rm jet}(\sqrt s,\Delta y,A)}{dp_T}\,,\\
\sigma_{\rm jet}\langle E_T\rangle(\psat,\sqrt s,\Delta y,A)
&=\int_{\psat} ^{\sqrt s/2}dp_T\,p_T\,
\frac{d\sigma_{\rm jet}(\sqrt s,\Delta y,A)}{dp_T}\,.
  \end{split}
  \]
The total number of minijets and the total amount of transverse
energy in $\Delta y$ in a nucleus-nucleus collision is obtained by
multiplying the corresponding nucleon-nucleon cross section with the
nucleon-nucleon luminosity of the collision (including an extra factor
of two for the number of minijets). This is given by the overlap
function $T_{AB}(b)$ of transverse densities $T_{A(B)}(s)$ of the
colliding nuclei:
  \[
  \begin{split}
T_{AB}(\bbb) &= \int d^2\bs T_A(|\bbb-\bs|)T_B(s)
=T_{AB}(b)\,,\\
T_A(\bs) &= \int_{-\infty}^{+\infty} dz\rho_A(z,\bs) = T_A(s)\,,
  \end{split}
  \]
where $\bbb$ is the impact parameter and $\bs$ the transverse
coordinate in nucleus $A$. For example the number of partons (which
can be defined only in lowest order because it is not an infrared-safe
quantity at higher orders) produced in a central zero-impact-parameter
collision of equal nuclei is
\be
\Delta N_{AA} = T_{AA}(0) \sigma_{\rm jet}(p_{\rm sat},\sqrt s,\Delta y,A)
\label{DeltaN}
\ee
in a rapidity interval $\Delta y$. A similar expression with
$\sigma_{\rm jet}$ replaced by $\sigma_{\rm jet}\langle E_T\rangle$
gives $\Delta E_T$, the transverse energy of minijets, in $\Delta y$.

Before discussing how to formulate the saturation condition to fix
$\psat$, we notice that the average densities are obtained by dividing
the total quantity with the volume that corresponds to the rapidity
interval $\Delta y$, $\Delta V= \Delta zA_T=\tau_0\Delta y\,\pi R_A^2$.
This procedure, with densities averaged over the transverse plane, is
easily generalized to local densities.  The nucleon--nucleon
luminosity in a transverse-area element $d^2\bs$ is
$T_A(|\bbb-\bs|)T_{B}(s)d^2\bs$, and the volume element is
$dV=\Delta z\,d^2\bs=\tau \Delta y\,d^2\bs$, leading to
\cite{Eskola:2001bf}
\be
n_{\rm pQCD}(\tau_0,\bs)={dN\over
\tau_0\Delta yd^2\bs}=\frac{2\sigma_{\rm
jet}}{\tau_0\Delta y} T_A(|\bbb-\bs|)T_{B}(s)
\label{ndensity}
%(p_0,\sqrt s,\Delta y,A)\,.
\ee
for the parton density and
\be
\epsilon_{\rm pQCD}(\tau_0,\bs)={dE_T\over\tau_0\Delta yd^2\bs}
=\frac{\sigma_{\rm jet}\langle E_T\rangle}
{\tau_0\Delta y} T_A(|\bbb-\bs|)T_{B}(s)
\label{edensity}
%(p_{\rm sat},\sqrt s,\Delta y,A)
\ee
for the energy density. The densities depend on the cut-off scale
through the cut-off dependence of cross sections.

The minijet cross sections above can be calculated separately for
gluon, quark, and anti-quark jets, allowing for the separate
determination of the densities of quarks and anti-quarks. From these
densities the initial net baryon number density is obtained as
$n_B=(n_q-n_{\bar q})/3$, which provides the initial condition for the
net-baryon-number current that satisfies the conservation law
(Equation \ref{eq:conscur}).

Up to here, we have essentially discussed how to obtain from a
boost-invariant momentum-space calculation of production cross
sections, $\sigma_{\rm jet}$ and $\sigma_{\rm jet}\langle E_T\rangle$,
the local densities $n_B$ and $\epsilon$. To close the calculation of
minijet cross sections, the saturation momentum $\psat$ must be
specified, and we do this by assuming that the parton (mainly gluon)
production saturates when the wave functions of produced partons start
to overlap.

\begin{figure}[htb]    % Example of Figure pull
\epsfxsize20pc         %
\centerline{\epsfbox{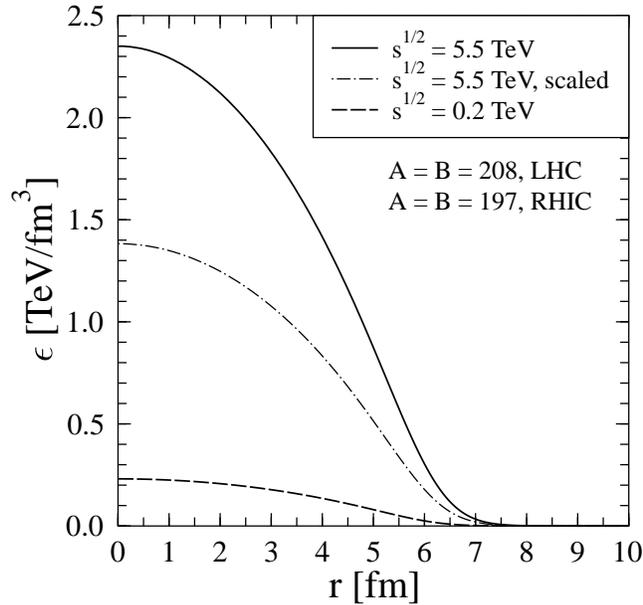}}
\caption{ \protect\small
Transverse dependence of the initial energy distribution for a
gold-on-gold collision at the Relativistic Heavy Ion Collider (RHIC)
(dashed line) and lead-on-lead collision at the Large Hadron Collider
(LHC) energy (solid line). The saturation scale is $p_{\rm sat} =1.16$
GeV at RHIC and 2.03 GeV at the LHC, with formation times of 0.170
fm/$c$ and 0.100 fm/$c$, respectively. The dashed-dotted line shows
the energy density at the LHC if $\tau_i=0.170$ fm/$c$ is used,
emphasizing the strong dependence of initial energy density on the
assumed initial time.}
\label{fig:inicond}
\end{figure}

In the transverse direction the scale of the wave functions is
$1/\psat$.  The scale in the longitudinal direction is not as obvious,
but we assume it is the same at the production time $\tau_0=1/\psat$.
At this time, the volume occupied by particles in the rapidity
interval $\Delta y$ equals $\Delta V=\tau_0\Delta yA_T$.  At
saturation, dividing this volume with the volume occupied by one jet,
$V_{\rm jet}$, should equal the number of produced partons $\Delta
N_{AA}$ in the rapidity interval $\Delta y$. This leads to the
condition
\be
\frac{\Delta N_{AA}(p_{\rm sat},\sqrt s,\Delta y,A)}{\Delta y}\,
\frac{\pi}{(\psat)^2} = A_T=\pi R_A^2.
\label{eq:saturation}
\ee
To avoid introducing a rather arbitrary cut-off parametrization at the
nuclear edges in transverse plane, we do not try to define the
calculation of $\psat$ locally in the transverse plane. Instead, an
effective value $\psat$, obtained from Equation~\ref{eq:saturation}
above, is used. Solving $\psat$ from this equation completes the
calculation of primary production in the pQCD+saturation model.  The
initial energy densities at $1/\psat$ at RHIC and LHC energies,
$\sqrt{s_{NN}}=200$ and $5500$ GeV, respectively, are shown in
Figure~\ref{fig:inicond}.

%%%%%%%%%%%%%%%%%%%%% FIGURE %%%%%%%%%%%%%%%%%%%%%%%%%%%%%%%%
%%%%%%%%%%%%%%%%%%%%%% FIGURE %%%%%%%%%%%%%%%%%%%%%%%%%%%%%%%%

The above formulation of calculating the initial densities is applied
in a central rapidity bin $|y|\leq \Delta y$. In the pQCD calculation,
particle production depends on the rapidity through the parton
distributions of colliding nuclei. However, when we use the results as
initial conditions for the hydrodynamic calculation, we assume that
boost-invariance is a good approximation at $y\approx0$ and take the
densities to give the boost-invariant initial conditions with no
$\eta$ dependence.

%% COLOR GLASS CONDENSATE; INITIAL STATE SATURATION

An attractive approach to particle production in heavy-ion collisions
at collider energies has been based on the assumption that the
initial-state parton densities saturate and non-linear dynamics
becomes dominant \cite{Gribov:tu}. With increasing collision energy,
the small-$x$ partons, in particular the gluons, become the dominant
part in the production, and their density in the initial wave function
of the nucleus becomes so high that gluons interact coherently and
their density saturates. The momentum below which gluons saturate is
called the saturation scale, $Q_{\rm s}$. It depends on the collision
energy $\sqrt s$ and the mass number of the nuclei.  From the point of
view of the color fields, the high density or large occupation numbers
of the field quanta with momenta $\lsim Q_{\rm s}$ can be described as
the formation of a color glass condensate.  This part dominates the
primary production at large $\sqrt s$ and can be treated in terms of a
classical effective field
theory~\cite{McLerran:1993ni,McLerran:1999hj}. Quantum corrections to
the classical treatment have also been
considered~(\cite{Kovchegov:1996ty,Jalilian-Marian:1996xn}; see also
Reference \cite{Kovchegov:2001sc} and references therein).

In the effective field theory approach to gluon production in $AA$
collisions, when boost-invariance is assumed, a gauge can be chosen
such that the problem can be formulated as a dimensionally reduced
2+1--dimensional theory.  A numerical approach~\cite{Krasnitz:1998ns}
with lattice regularization can be applied to the reduced theory.

The numerical calculations with lattice regularization started using
SU(2) symmetry \cite{Krasnitz:1999wc,Krasnitz:2000gz} and other
simplifications, such as cylindrical nuclei, but were soon formulated
for SU(3) \cite{Krasnitz:2001qu,Krasnitz:2002mn} and realistic nuclear
geometry.  Also, local color neutrality in the transverse overlap
region of the collision \cite{Krasnitz:2002mn} was imposed and other
smaller inconsistencies corrected, leading to a ratio of $E_{_T}/N$,
which is consistent with the physical interpretation of the saturation
scale \cite{Lappi:2003bi}.

The lattice approach does not, however, give a value of the saturation
scale itself; the overall normalization must be obtained from
elsewhere.  For RHIC phenomenology, the authors of Reference
\cite{Krasnitz:2002mn} suggest two sets of results, which, in light of
the latest results of Reference \cite{Lappi:2003bi} lead to following,
qualitatively different descriptions of the final state:
%%%
\begin{itemize}
\item
In the case of a smaller scale $\mu$, the total transverse energy
$E_{_T}\sim g_s^4R_A^2\mu^3$ produced from the classical fields
roughly equals the experimentally measured result, whereas the number
of initially produced partons ($N\sim g_s^2R_A^2\mu^2$) is only about
half of the multiplicity of hadrons measured in the experiment. In
this case, the only change in the final state is the fragmentation of
partons to $\sim 2$ hadrons on average. However, there would be no
significant hydrodynamic evolution because this would reduce the
transverse energy below the measured value. In this picture, which
corresponds to the scenario suggested by Kharzeev \& Levin
\cite{Kharzeev:2001gp}, one would expect the photon and lepton pair
emission after the primary interactions to be very rare.
\item
For a larger saturation scale, the number of partons is close to the
measured number of hadrons but the initially produced transverse
energy is approximately 2.5 times bigger than the measured one
\cite{Lappi:2003bi}. In this case, production must be followed by
strong initial collective expansion in the longitudinal direction,
allowing for a transfer of energy into the longitudinal motion.  This
case corresponds to the evolution suggested by pQCD + saturation +
hydrodynamics calculation \cite{Eskola:2001bf}.
\end{itemize}

The energy dependence in the above models enters through the
dependence of the saturation scales on the center-of-mass energy
leading, for example, to rather similar growth of multiplicity from
RHIC to the LHC in both models.

%%_RADIAL_DEPENDENCE_OF_DENSITIES:_WOUNDED_NUCLEON_MODEL__

For a head-on, zero-impact-parameter collisions, the produced system
is cylindrically symmetric, all quantities depend only on $\tau$ and
$r$, and the transverse flow is radial with no azimuthal
dependence. We show results on hadron spectra displaying effects from
radial flow. However, a good test for the applicability of the
hydrodynamic description of heavy ion collisions is provided by
nonzero-impact-parameter collisions without cylindrical symmetry. In
our example of the calculation of primary densities the expressions
(\ref{ndensity}, \ref{edensity}) hold also for
nonzero-impact-parameter collisions. However, the determination of
saturation scale becomes more involved, and a straightforward
generalization of the saturation condition (Equation
\ref{eq:saturation}) leads the multiplicity to have a too-flat
dependence on the number of participant nucleons, a possible measure
of non-centrality of the collision~\cite{Kolb:2001}.

Above, the transverse dependence of the initial densities is given by
the number of collisions per unit transverse area:
$n_{\rm{coll}}(\bbb,\bs)~\propto~T_A(|\bbb-\bs|)T_{B}(s)$.  In another
popular phenomenological approach, the initial densities are assumed
to be proportional to the number of participants, also known as
wounded nucleons\footnote{Strictly speaking wounded nucleons are
nucleons that scatter inelastically whereas participants are nucleons
that scatter elastically or inelastically. In the recent literature
this difference is usually ignored and both terms are used in the
sense of participants.} per unit transverse area. In the eikonal
Glauber model, this is defined as
\bea  %{init}
   n_{\rm WN}(\bs;\bb) 
    & = &  T_A(\bs+\mhalf\bb) \displaystyle{
           \Bigl[1-\Bigl(1-\frac{\sigma T_B\bigl(\bs-\mhalf\bb\bigr)}
                                {B} \Bigr)^B \Bigr]   }%end displaystyle
 \nonumber\\
    & + &  T_B\bigl(\bs-\mhalf\bb\bigr) \displaystyle{
           \Bigl[1-\Bigl(1-\frac{\sigma T_A\bigl(\bs+\mhalf\bb)\bigr)}
                                {A}\Bigr)^A \Bigr]\,, }
\eea
where $\sigma$ is the nucleon-nucleon cross section. For a
zero-impact-parameter collision, this indicates, except at the edges
of the nuclei, a radial dependence approximately proportional to the
sum of thickness functions $T_A(r)$ and $T_B(r)$. In the central
region of the overlap area, the resulting density distribution is
flatter than when the density is proportional to the number of
collisions. This means smaller pressure gradients and slower evolution
of transverse flow.

In the literature, both the proportionality to the number of binary
collisions and to the number of participants has been used to fix
either the initial entropy density or energy density. The
proportionality constant is chosen to reproduce the measured final
particle multiplicity in most central collisions. The centrality
dependence of the multiplicity is then predicted by the model, but at
RHIC, neither binary-collision scaling nor wounded-nucleon scaling
reproduces the data. However, a linear combination of them does and is
therefore used to describe the initial density
distribution~\cite{Heinz,Heinz2}.

%% INITIAL TRANSVERSE VELOCITY
We have not yet specified the initial transverse velocity $\bv_T$.
Usually this is taken to be zero. This choice is supported by the
argument that the final state in each primary collision is randomly
oriented in the transverse plane and thus one expects the
transverse-momentum average in any volume element to vanish.  There is
a slight flaw in this argument because the transverse density of
produced partons is not constant, and this can lead to a nonzero
momentum average in a (finite) volume element.  However, comparison to
experimental data shows that for agreement, only small initial
transverse velocities are allowed. In the calculations shown below,
$\bv_T(\tau_0,\br)=0$ has been used.

%% THERMALIZATION

Because the use of hydrodynamics presumes thermal equilibrium, the
time scale for thermalization after the primary production must be
fixed.  The dynamics of thermalization can be even more difficult to
solve than that of primary production. Results of theoretical studies
of thermalization have not yet converged. In the so-called bottom-up
thermalization scenario~\cite{Baier:2000sb}, thermalization times are
predicted to be long, of the order of 2\ldots 4~fm/$c$. A more recent
idea of the role of instabilities in the expansion predicts much
shorter thermalization times of order below 1
fm/$c$~\cite{Mrowczynski:2005ki}.

The thermalization time scale is an important issue because a
hydrodynamic description of the elliptic flow can be achieved only if
the thermalization time is short. In the numerical examples we provide
below, we always use thermalization times below 1~fm/$c$. When showing
results based on the minijet initial state, we use the production time
scale as the thermalization time, $\tau_i=1/\psat$.

%%%%%%% EoS %%%%%%%%%%%

\subsection{The Equation of State of Strongly Interacting Matter}
\label{sec:EoS}

A major complication in the description of the evolution of the matter
produced in a high-energy collision of nuclei is the change in the
degrees of freedom. The dense initial-state parton matter expands and
turns into a gas of hadrons and hadron resonances when dilute and cool
enough. According to the present understanding, at large $\mu_B$ and
small or moderate $T$, there exist different correlated phases, such
as the phase with color-flavor locking~\cite{Rajagopal:2000uu}.
Nearing smaller $\mu_B$, the transition between hadron resonance gas
and QGP is believed to be of first order when $\mu_B$ is not too
small.  For two light quarks and one heavy quark, the phase boundary
is conjectured to end at a critical point, and below that the
transition is a rapid cross-over. The quantitative theoretical
information from QCD lattice simulations, which have recently been
extended from the $T$--axis to finite chemical potential, supports
this picture.  There are also arguments in favor of the existence of
strong correlations in the quark matter close to the phase
boundary. These may explain the ideal-fluid behavior of QGP indicated
by successful hydrodynamical explanation of elliptic flow. When matter
is assumed to be in the state of non-interacting quarks and gluons,
the ideal QGP, a simple ideal gas EoS of massless particles
($P=\epsilon/3$) is often used to describe it. A more sophisticated
but less usual way is to use parametrized lattice QCD results.

In heavy ion collisions at collider energies the net-baryon number is
small, with $\mu_B\lsim 50$~MeV~\cite{PBM}, indicating a cross-over
transition. Somewhat unexpectedly, from the point of view of
hydrodynamic expansion, the difference between a weak first-order
transition and a rapid cross-over is not very significant, as long as
the EoSs are relatively similar away from the transition region and
the increase in entropy and energy densities around the critical
temperature is sufficiently large and rapid~\cite{Huovinen-EoS}. Upon
a closer look this is not so surprising because the main qualitative
feature is a jump in the thermodynamical densities $\epsilon$ and
$s$. The size of the jump depends essentially on the size of the
change of the number of degrees of freedom.  From the point of view of
hydrodynamics, the rapid jump in $\epsilon$ combined with a slower
change in pressure appears as a softening of the EoS. It is seen as a
slow-down in the acceleration of the transverse flow in the transition
region. Details differ for the two cases, but the final features of
flow are quite similar and the quantitative differences in the final
hadron spectra are small.

An interacting hadron gas can be described in good approximation as a
gas of noninteracting hadrons and resonances. The inclusion of
resonances mimics the effects of both attractive and repulsive
interactions between hadrons reasonably well~\cite{Raju}. However, the
repulsive interaction between baryons at large net-baryon densities
must be included as an additional mean field~\cite{Kapusta:1982qd} or
as an excluded volume correction~\cite{Rischke:1991ke} to give a
reasonable phase-transition behavior between hadronic and partonic
phases. A detailed account of constructing an EoS with a mean field
can be found, for example in Reference~\cite{Sollfrank_prc}.

In calculations, we use an EoS with ideal QGP in the high-temperature
phase and a hadron resonance gas with a mean field below the
transition. A first-order transition is implemented by introducing a
bag constant $B$ into the QGP phase and connecting the two phases with
a Maxwell construction. We use $N_f=3$, and the bag constant $B$ and
mean field constant $K$ are chosen to be $B^{1/4}=243$ MeV and $K=450$
MeV fm$^3$, respectively, giving $T_c=167$ MeV for the transition
temperature.

An additional complication in constructing an EoS of a hadron gas
relevant for heavy ion collisions is the chemical composition of the
hadron gas.  The usual assumption of hydrodynamics is that of chemical
equilibrium.  This assumption is supported by thermal models that can
reproduce the observed hadron abundances by assuming a thermal source
in $T\approx 170$ MeV temperature.  However, many studies have found
that the $p _T$ distributions of hadrons are better described by
assuming a colder, flowing source in $T=100$--140 MeV temperature.
Thus, the assumption of chemical equilibrium between these
temperatures is questionable. In many hydrodynamical calculations,
this observation is simply ignored and chemical equilibrium is assumed
to hold until kinetic freeze-out at $T=100$--140 MeV. In such cases
one can reproduce the slopes of the hadron $p_T$ spectra, but it is
not possible to reproduce simultaneously both baryon and antibaryon
yields.

One solution to this problem is the so-called single freeze-out
model~\cite{Florkowski}, in which a suitable choice of freeze-out
surface allows one to fit the $p_T$ spectra, even if the kinetic
freeze-out temperature is taken to be the same $T\approx 165$ MeV as
the chemical freeze-out temperature. As we show below, in the context
of hydrodynamical models, similar approach with $T\approx 150$ MeV can
be used to reproduce, at least approximately, the hadron $p_T$
distributions in most central collisions at RHIC
energies~\cite{Eskola:2002wx}. Whether the anisotropies of particle
distributions (see section \ref{elliptic}) can also be reproduced this
way, has not been tested so far.

Another solution to this problem is to assume two separate freeze-outs
--- chemical and kinetic --- and to modify the EoS between these
temperatures accordingly. In such an approach, the hadron yields are
assumed to be fixed at some chemical freeze-out temperature, usually
soon below or at the phase-transition temperature. These hadron yields
are subsequently described as conserved currents, and each conserved
hadron species is assigned a chemical potential. This way the yields
of all hadron species can be reproduced separately, even if a low
kinetic freeze-out temperature is used~\cite{Bebie:1991ij}.

Such an EoS changes the buildup of collective motion ---~i.e.~flow~---
very little because pressure as a function of energy density,
$P = P(\epsilon)$, is very similar to the chemical equilibrium
EoS~\cite{Teaney-EoS,Hirano2}. However, temperature as a function of
energy density changes radically, and when collective and thermal
motion are folded to calculate observable particle distributions, the
results differ~\cite{Hirano2,Peter-Ralf}.

%%%%%%%%%%%%%%%%%%%%%%%%%%%%%%%%%%%%%%%%%%%%%%%%%%%%%%%%%%%%%%%%%%%%%
\subsection{Transverse flow}
\label{sec:transverse_flow}
%%%%%%%%%%%%%%%%%%%%%%%%%%%%%%%%%%%%%%%%%%%%%%%%%%%%%%%%%%%%%%%%%%%%%

\begin{figure}[htb]    % Example of Figure pull
\epsfxsize20pc         %
\centerline{\epsfbox{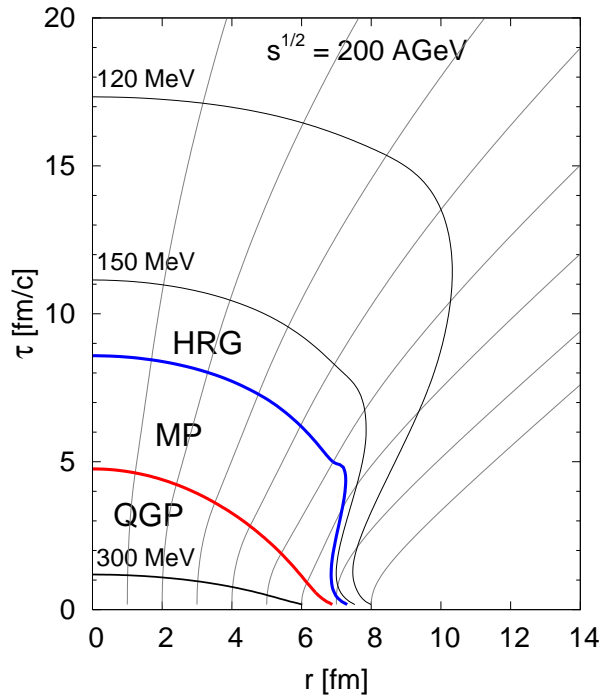}}
\caption{ \protect\small
 Temperature contours at 300 (in quark-gluon plasma QGP), 150 and 120~MeV
 (hadron resonance gas, HRG), and the boundaries of mixed phase (MP) with
 QGP and HRG at $T_c=167$ MeV. Flow lines are also shown. Initial
 conditions are from a pQCD~+ saturation calculation at
 $\sqrt{s_{NN}}=200$ GeV.  Note that the slope of the flow line is
 related to the velocity and the curvature to the acceleration.}
\label{fig:flowlines200}
\end{figure}

To illustrate the transverse flow, Figure~\ref{fig:flowlines200} shows
the boundaries of QGP and hadron gas, with the mixed phase between
them. Also, three contours of temperature are depicted, as well as the
flow lines with 1-fm intervals starting at $\tau_0$. The initial
conditions are those from the pQCD + saturation model at RHIC energy
$\sqrt{s_{NN}}=200$ GeV.
%
%%%%%%%%%%%%%%%%%%%%% FIGURE %%%%%%%%%%%%%%%%%%%%%%%%%%%%%%%%
%%%%%%%%%%%%%%%%%%%%%% FIGURE %%%%%%%%%%%%%%%%%%%%%%%%%%%%%%%%

According to this calculation, the maximum lifetime of the plasma
phase is $5$~fm/$c$, and that of the mixed phase is $\sim 8$~fm/$c$.
The flow lines show how the fluid elements accelerate and move. The
slope of the flow line is related to the velocity of the fluid and the
curvature to the acceleration.  In the mixed phase, where the pressure
is constant, flow lines are straight because there is no acceleration.
At small $r$, gradients are small and the flow lines bend slowly. In
particular, in the plasma, when $r$ increases, the pressure gradient,
and consequently the acceleration, increases, indicated by the faster
bending of flow lines. Along the flow line starting at $r=8$~fm/$c$,
the densities are small even at $\tau_0$, and this region is
insignificant in calculating the spectra. For calculational reasons
the initial densities are taken to go smoothly to zero, and the
hydrodynamic equations are also solved at large values of $r$.

\section{HADRON DISTRIBUTIONS AND CORRELATIONS}
  \label{hadrons}

%%%%%%%%%%%%%%%%%%%%%%%%%%%%%%%%%%%%%%%%%%%%%%%%%%%%%%%%%%%%%%%%%%%%%
\subsection{Freeze-out and the calculation of hadron spectra}
\label{sec:spectra_hadron}
%%%%%%%%%%%%%%%%%%%%%%%%%%%%%%%%%%%%%%%%%%%%%%%%%%%%%%%%%%%%%%%%%%%%%

As matter expands, distances between particles become large,
collisions cease and momentum distributions freeze out. The condition
for the freeze-out to occur is usually expressed locally in terms of
the energy density or temperature reaching a given value. This
determines a three-dimensional freeze-out surface $\sigma^\mu(x)$ in
space-time. The prescription of Cooper and Frye \cite{Cooper:1974mv}
convolutes the flow with the thermal motion along the freeze-out
surface:
\bea
 \pi{dN\over d^3{\bf p}\slash E}
&=& {dN\over dydp_{_T}^2} = \pi\int_\sigma
d\sigma_{\mu}(x)p^{\mu}f(x,p;T(x)) \\
&=& {g\over 2\pi}
\sum_{n=1}^\infty (\pm 1)^{n+1} \int_\sigma r\tau \big[-p_{_T} I_1(n
\gamma_rv_r{p_{_T}\over T}) K_0(n\gamma_r {m_T\over T})\,d\tau
\nonumber\\
& & +m_T I_0(n \gamma_rv_r{p_{_T}\over T})
K_1(n\gamma_r{m_T\over T})\,dr
\big]\,.
\label{eq:CooperFrye}
\eea
The second expression above is valid for cylindrically symmetric,
boost-invariant flow with $v_r$ the radial flow velocity,
$\gamma_r=1/\sqrt{1-v_r^2}$, and $K_i$ and $I_i$ are Bessel functions
of second kind.

%%%%%%%%%%%%%%%%%%%%%%_1_FIGURE_%%%%%%%%%%%%%%%%%%%%%%%%%%%%%%%%
\begin{figure}[htb]    % EXAMPLE OF FIGURE PULL   !!!!!!!
\epsfxsize18pc         %
\centerline{\epsfbox{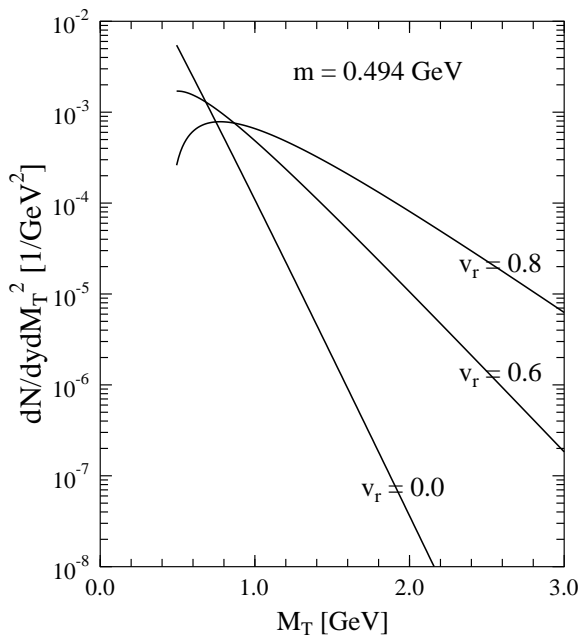}}
\caption{\protect\small
The effect of flow on the spectrum of kaons. Temperature is
kept unchanged and the spectrum is shown for radially flowing
matter at velocities $v_r=0,\,\, 0.6$ and $0.8$.}
\label{fig:kaon+flow}
\end{figure}
%%%%%%%%%%%%%%%%%%%%%%_1_FIGURE_%%%%%%%%%%%%%%%%%%%%%%%%%%%%%%%%

The unstable particles are treated as follows: First, the spectra of
all hadrons and hadron resonances are calculated using Equation
\ref{eq:CooperFrye}. We then follow the chains of all possible two-
and three-body decays and collect the spectra of final stable
hadrons~\cite{Sollfrank:1991}. Stable hadrons can either be
interpreted as all those that do not decay through strong
interactions, or we can follow the feed-downs via weak and
electromagnetic decays. E.g.\ we can calculate the $\pi^0$ spectrum
including both the $\pi^0$'s at freeze-out and all the decays that
lead to $\pi^0$'s, and then study the photon spectrum from $\pi^0$
decays alone or from all decays with photons in the final
state. Phenomenologically an important case is that of feed-down
nucleons from weak decays of strange hyperons. Here again, we can
calculate, for example, the spectrum of $\Lambda$'s and then see how
it contributes through weak decays to the spectrum of protons.

Before showing our results at RHIC and LHC energies we discuss how the
radial flow affects a spectrum. In Figure~\ref{fig:kaon+flow} we show
the spectrum of kaons from matter at rest or flowing at velocities
$v_r=0.6$ and 0.8. We assume that the matter decouples at a fixed time
so that only the second term in Equation \ref{eq:CooperFrye}
contributes. The temperature of the matter is the same in each case
and is essentially the inverse of the logarithmic slope of the
spectrum for $v_r=0$. From the asymptotic properties of modified
Bessel functions of second kind, it is clear that at large $m_T$ the
change in the slope, when $v_r$ changes to a nonzero value, can be
expressed by replacing the temperature $T$ with an effective
temperature $T_{\rm{eff}}=T\sqrt{(1+v_r)/(1-v_r)}$. For $v_r=0.6$ the
change is a factor of two and for $v_r=0.8$ a factor of three.

%%%
%%%%%%%%%%%%%%%%%%%%%%%%%%%%%%%%%%%%%%%%%%%%%%%%%%%%%%%%%%%%%%%%%%%%%
\subsection{Transverse momentum spectra of hadrons}

%%%%%%%%%%%%%%%%%%%%%%%%%%%%%%%%%%%%%%%%%%%%%%%%%%%%%%%%%%%%%%%%%%%%%
%%%

Next we compare some of the hydrodynamical results
\cite{Eskola:2005ue} with the experimental transverse momentum spectra
measured by the STAR \cite{Adams:2003kv, Adler:2002xw, Adams:2003xp,
Adler:2002uv}, PHENIX \cite{Adler:2003au, Adcox:2001jp, Adcox:2001mf,
Adcox:2002au, Adler:2003cb, Adler:2003qi}, PHOBOS \cite{Back:2003qr,
unknown:2004zx} and BRAHMS \cite{Arsene:2003yk, Bearden:2004yx,
Bearden:2003hx} collaborations for the most central bins in Au+Au
collisions at $\ssNN=130$ and 200 GeV. The calculated spectra are
obtained by using Equation \ref{eq:CooperFrye} with the flow
illustrated in Figure~\ref{fig:flowlines200}.

Note that a hydrodynamic calculation cannot describe the hadron
spectra at large transverse momenta. At large $p_T$ the hydrodynamical
calculation shows an approximate exponential behavior, whereas the
tails of measured spectra essentially obey a power law. At RHIC, the
transition from steep exponential to a shallower power behavior takes
place at $p_T\sim 3$ GeV. The fraction of hadrons with $p_{_T}\gsim3$
GeV from all hadrons is small, and they originate from the
fragmentation of high-energy partons, which suffer some energy loss in
the dense medium of low-energy partons, but are not thermalized. We
return to the interplay of the low-energy partons, which provide the
main transverse energy and are assumed to thermalize, and the
high-energy partons, which lose some fraction of energy in
rescattering but require a much larger system for thermalization.

\begin{figure}[!ht]
 \vspace{-0.7cm}
    \hspace{-0.9cm}
    \epsfxsize 85mm \epsfbox{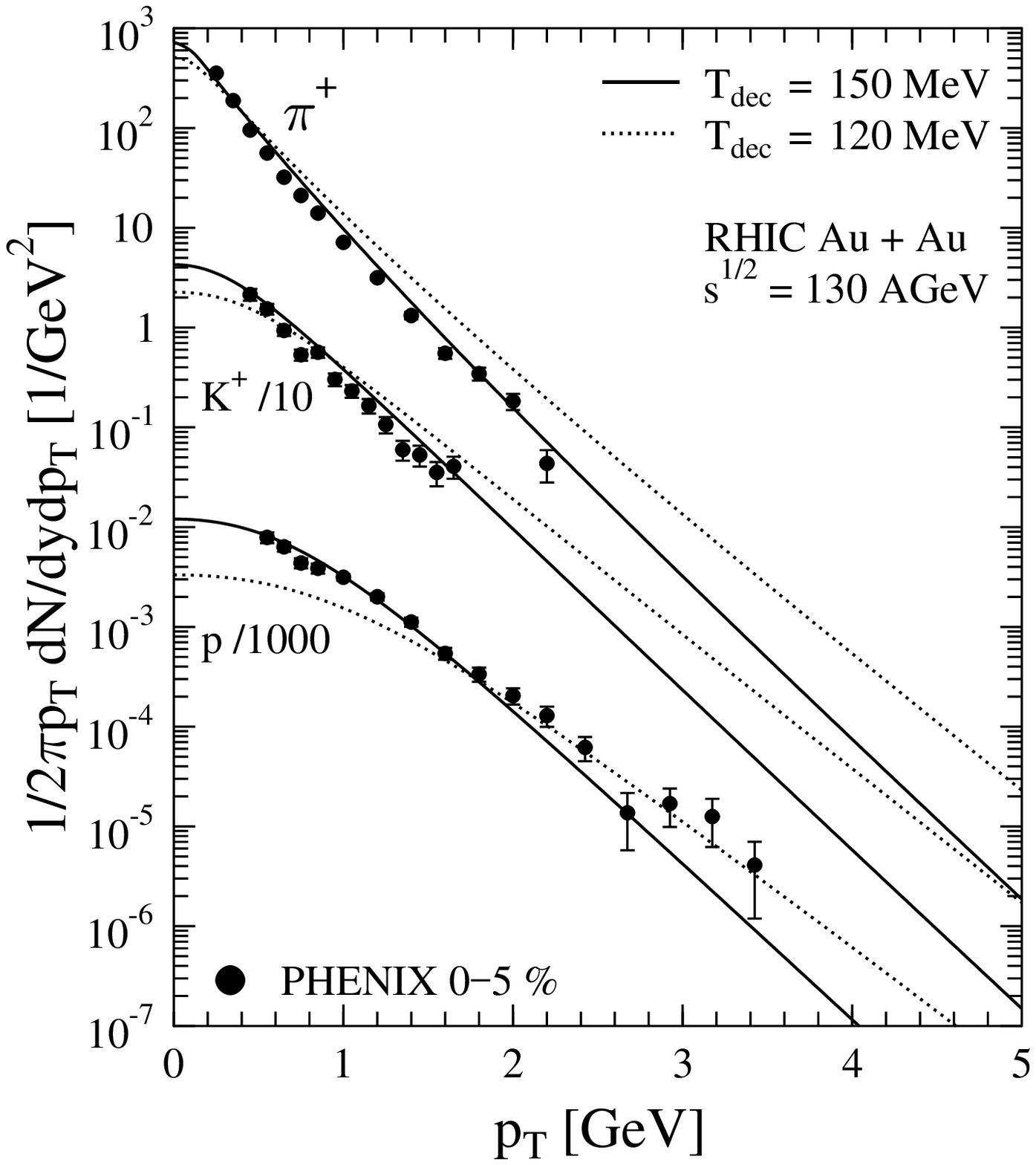}  \hspace{-1.8cm}
    \epsfxsize 85mm \epsfbox{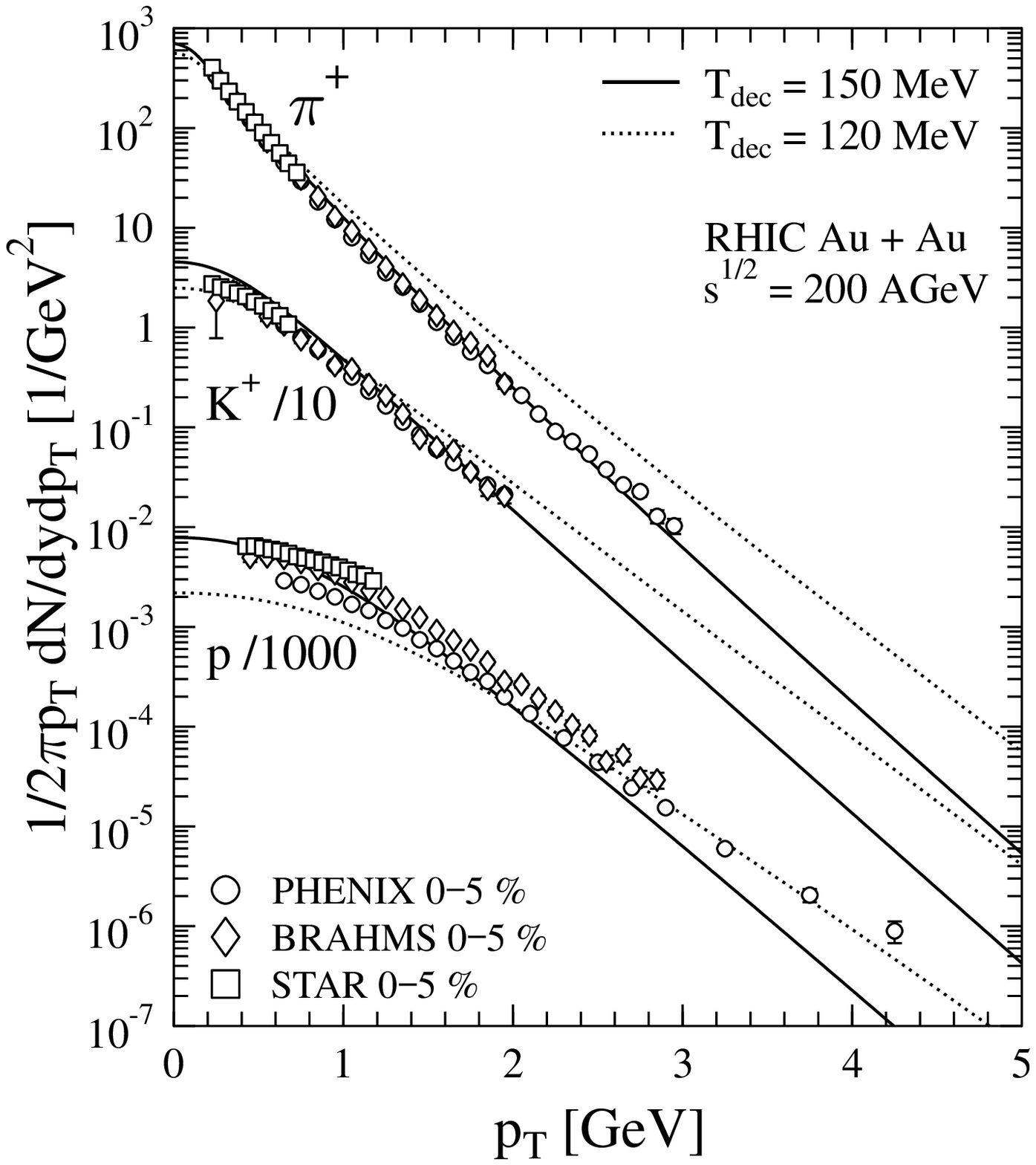}

 \vspace{-2.5cm}
\begin{minipage}[t]{68mm}
 \begin{center}
  \caption{\protect\small
Transverse momentum spectra of positive pions, positive kaons, and
protons at $y=0$ in the most central 5 \% of Au+Au collisions at
$\ssNN=130$~GeV. The solid and dotted lines show our hydrodynamic
results for freeze-out temperatures $\Tdec=150$~MeV and
$\Tdec=120$~MeV, respectively. The PHENIX data \cite{Adcox:2001mf} is
plotted with the given total error bars. Note the scaling factors 10
and 1000 for kaons and protons, respectively. Both the hydrodynamic
result and the PHENIX data contain the feed-down contributions from
hyperons.
}
  \label{fig:positive130}
 \end{center}
\end{minipage}
\hspace{\fill}
\begin{minipage}[t]{62mm}
 \begin{center}
  \caption{\protect\small
As Figure~\ref{fig:positive130} but at $\ssNN=200$~GeV. The PHENIX
data~\cite{Adler:2003cb} and the BRAHMS data~\cite{Bearden:2004yx,
Bearden:2003hx} are shown with statistical errors and the STAR
data~\cite{Adams:2003xp} with the given total error bars. The
hydrodynamic calculation and the PHENIX data are without the hyperon
feed-down contributions whereas the STAR and BRAHMS data contain the
feed-down.
}
  \label{fig:positive200}
 \end{center}
\end{minipage}
\end{figure}

We start with the $p_T$ spectra of identified hadrons at
midrapidities. Figure~\ref{fig:positive130} shows the PHENIX data
collected for positive pions, kaons and protons in the most central 5 \%
of Au+Au collisions for $y=0$ at $\ssNN=130$~GeV~\cite{Adcox:2001mf}.
Similarly, in Figure~\ref{fig:positive200} STAR~\cite{Adams:2003xp},
PHENIX~\cite{Adler:2003cb} and BRAHMS~\cite{Bearden:2004yx,Bearden:2003hx}
data are shown at $\ssNN=200$~GeV. Note the scaling factors 10 and
1000 for kaons and protons, respectively. An important issue of
uncertainty in the calculation is the dependence of the results on the
decoupling temperature. This is shown by plotting the results for
freeze-out temperatures $T_{\rm dec}=150$ MeV (solid lines) and 120
MeV (dotted lines). Note that the normalization of the pion spectrum
is almost independent of the decoupling temperature. Because pions
provide the main contribution to the total multiplicity, $dN_{\rm
tot}/dy$ depends only weakly on $T_{\rm dec}$. However, the
multiplicity of heavier particles is very sensitive to $T_{\rm dec}$,
as expected. The shape of all three spectra changes clearly with the
decoupling temperature.  This follows from the increase in flow
velocity, characterized by the increase of the effect with the
increasing mass of the particle. The aim of this calculation has not
been to find a best fit to the data, but the results show that both
the normalization and the slope of the data at momenta $p_{_T}\lsim3$
GeV can be described quite well with a single $T_{\rm dec}$ in the
neighborhood of 150 MeV.

%%%%%%%%%%%%%%%%%%%%% BEGIN FIGURE %%%%%%%%%%%%%%%%%%%%%%%%%%%%%%%%
%%%%%%%%%%%%%%%%%%%%% END FIGURE %%%%%%%%%%%%%%%%%%%%%%%%%%%%%%%%

\begin{figure}[!ht]
    \vspace{-1.7cm}
    \hspace{-0.9cm}
    \epsfxsize 85mm \epsfbox{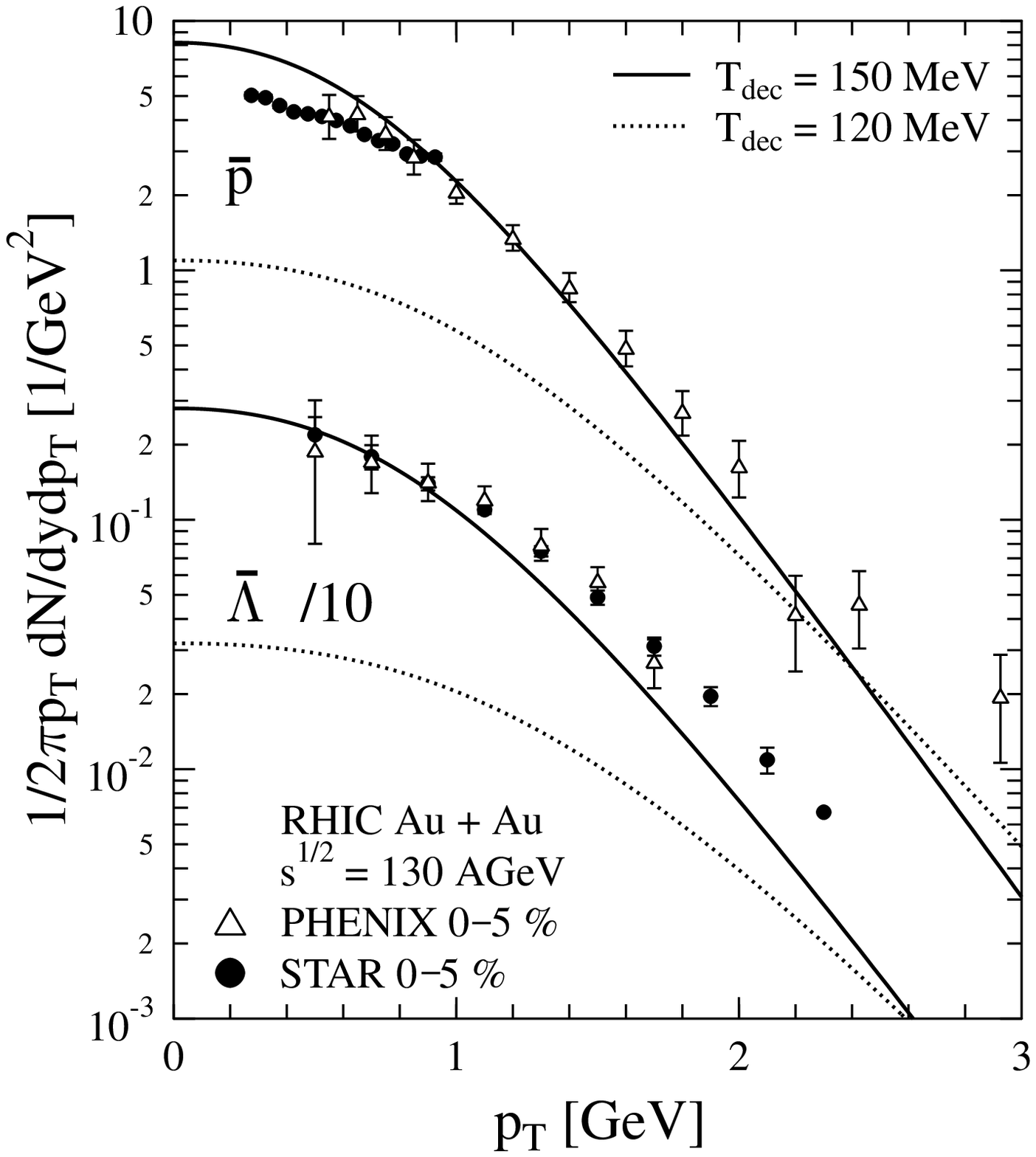}  \hspace{-1.8cm}
    \epsfxsize 85mm \epsfbox{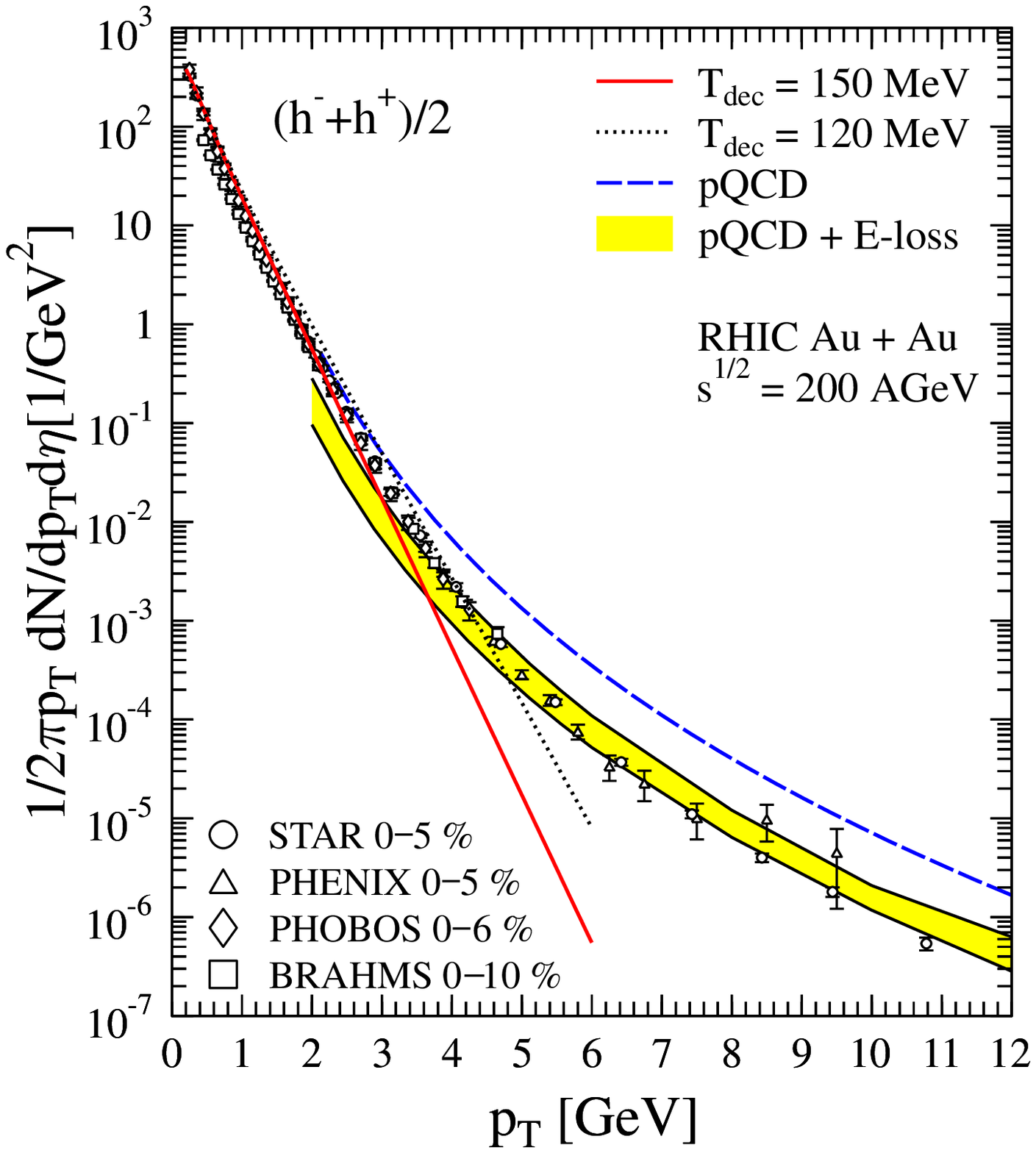}

  \vspace{-1.5cm}
\begin{minipage}[t]{64mm}
%    \vspace*{-1.2cm}
    \begin{center}
    \caption{\protect\small
Transverse momentum spectra of antiprotons and antilambdas at $y=0$ in
the most central 5 \% of Au+Au collisions at $\ssNN=130$~GeV. Our
hydrodynamic results are for freeze-out temperatures $\Tdec=150$~MeV
(solid line) and $\Tdec=120$~MeV (dotted line), with hyperon feed-down
contributions included, as in the PHENIX $\bar p$ \cite{Adcox:2001mf}
and $\bar\Lambda$ \cite{Adcox:2002au} data and the STAR
data~\cite{Adler:2002uv}.
}
    \label{fig:lambda130}
    \end{center}
\end{minipage}
\hspace{\fill}
\begin{minipage}[t]{64mm}
%    \vspace*{-1.2cm}
    \begin{center}
    \caption{\protect\small
Transverse momentum spectra of charged particles at $\eta=0$ (averaged
over $|\eta|\le0.1$) in the most central 5 \% of Au+Au collisions at
$\ssNN=200$~GeV. Our hydrodynamic results are shown for
$\Tdec=150$~MeV (solid line) and $\Tdec=120$~MeV (dotted line). The
pQCD fragmentation results are shown with (shaded band) and without
(dashed line, see the text) energy losses.  The data is taken by
STAR~\cite{Adams:2003kv}, PHENIX~\cite{Adler:2003au},
PHOBOS~\cite{Back:2003qr}, and BRAHMS~\cite{Arsene:2003yk}.
}
    \label{fig:charged200}
    \end{center}
\end{minipage}
\end{figure}

For the identified particles in Figures~\ref{fig:positive130}
and~\ref{fig:positive200}, the measured spectra do not extend to large
enough $p_T$ to show clearly the deviation from the hydrodynamic
results, with the exception of proton spectra. If the decoupling
temperature is 150 MeV to reproduce the normalization, the slope tends
to be too steep. The proton yield from jet fragmentation, as explained
in detail in Reference~\cite{Eskola:2005ue}, does not seem to be large
enough in the $p_T\sim 3\ldots 5$ GeV region to bring the calculation
into agreement with the data. In Figure~\ref{fig:lambda130}, the
spectra of two other heavy particles are shown, those of antilambdas
and antiprotons. These show the same trend as protons, pointing to the
need for separate chemical and kinetic decoupling when describing
simultaneously the details of all spectra. Studies with separate
chemical and kinetic decoupling, in which the stable particle numbers
are fixed after chemical freeze-out, indeed show that the spectra of
pions and kaons become almost independent of the kinetic decoupling
temperature $T_{\rm dec,kin}$, whereas the (anti)proton spectra widen
with decreasing $T_{\rm dec,kin}$~\cite{Hirano2,Peter-Ralf}. There
are, however, claims in the literature that separate chemical and
kinetic decoupling lead to a worse overall fit to the slopes of $p_T$
distributions than what can be achieved by requiring chemical
equilibrium until kinetic
freeze-out~\cite{Hirano:2005wx,Hirano:2005xf}. Studies exploring the
effects of initial time, the shape of initial distributions, and the
value of $T_{\rm dec,chem}$ while using two separate freeze-outs are
needed to settle the issue.

%%%%%%%%%%%%%%%%%%%%% BEGIN FIGURE %%%%%%%%%%%%%%%%%%%%%%%%%%%%%%%%
%%%%%%%%%%%%%%%%%%%%% END FIGURE %%%%%%%%%%%%%%%%%%%%%%%%%%%%%%%%

The range where hydrodynamics can be used to describe the hadron
spectra is indicated clearly in Figure~\ref{fig:charged200}, which
shows results from our hydrodynamical calculations and from a pQCD jet
calculation, followed by an energy loss in the medium before the jet
fragments into hadrons (see below). The STAR and PHOBOS data are
plotted with the given total error bars, the PHENIX data by adding the
given statistical and systematic errors in quadrature, and the BRAHMS
data with the given statistical error bars.

The transverse spectrum up to $\sim 3$~GeV is similar to that of the
dominant pion component shown separately in
Figure~\ref{fig:positive200}.  It has the shape typical of the
spectrum from hydrodynamic calculations, falling off roughly
exponentially. In the region $p_T\sim3\ldots4$ GeV, there is a large
change in the slope, indicating a change in the overall production
mechanism. The calculation of primary production proceeds through the
hard and semihard interactions between the partons of the incoming
nuclei, both in the case of initial conditions for hydrodynamical
equations and the energy loss of the jets. In the hydrodynamic
calculation, the produced partons are assumed to thermalize quickly
and then undergo hydrodynamic expansion in local thermal equilibrium
until the freeze-out. In the calculation of jet fragmentation after
energy loss, the produced \emph{high-energy} partons are assumed to
survive the thermalization, but lose energy when traversing the
thermal medium formed by the lower-energy partons.  When energy loss
and fragmentation are taken into account, the original energy of the
partons that fragment to hadrons of $p_T\gsim3$ GeV must be of the
order of $\sim 6$~GeV or greater. It turns out that the contribution
of partons with $p_T\gsim4$ GeV to the production of (transverse)
energy is less than 5 \%, justifying as a good approximation the
assumption that all produced transverse energy is thermalized. The
details of the jet-energy-loss and fragmentation calculation are
explained in References~\cite{Eskola:2005ue} and \cite{Eskola:2004cr}.

%%%%%%%%%%%%%%%%%%%%%% BEGIN FIGURE %%%%%%%%%%%%%%%%%%%%%%%%%%%%%%%%
\begin{figure}[tbh]
     \vspace{-1.6cm}
    \epsfxsize20pc
    \centerline{\epsfbox{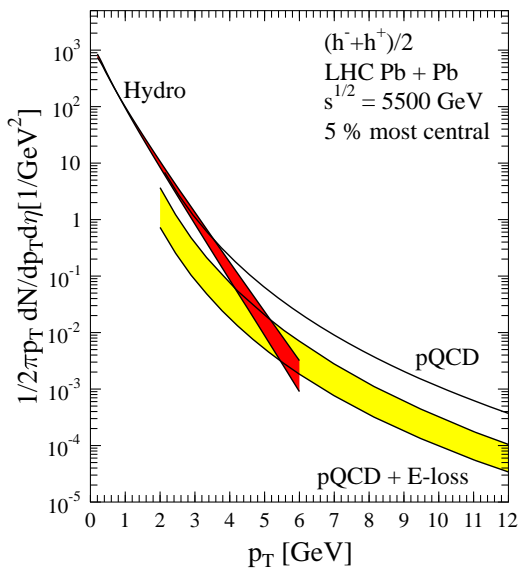}}
     \vspace{-1.2cm}
\caption{\protect\small
Predictions for the transverse momentum spectra of charged hadrons at
$\eta =0$ in the most central 5 \% of Pb+Pb collisions at Large Hadron
Collider (LHC) energy $\ssNN=5500$~GeV. The shaded band in the
hydrodynamic results shows the freeze-out temperature interval
$120\dots 150$~MeV. The solid curve labeled "pQCD" corresponds to the
pQCD fragmentation results without energy losses. The shaded band
labeled "pQCD + E-loss" describes the uncertainty in the pQCD
fragmentation results with energy losses. }
  \label{fig:chargedLHC}
\end{figure}
%%%%%%%%%%%%%%%%%%%% end figure %%%%%%%%%%%%%%%%%%%%%%%%%%%

Although the two-component approach, hydrodynamically expanding
thermal matter as the source of low-$p_T$ hadrons and jet
fragmentation after energy loss producing the high-$p_T$ hadrons,
seems reasonable and justified, adding them straightforwardly would be
too naive. In the region of turn-over from one mechanism to the other,
both contribute, but part of the hadrons cannot be assigned to either
component. Other mechanisms like recombination can also contribute in
this region~\cite{Fries:2004ej}. Accurate data in this area would be
useful in understanding both the energy loss and thermalization of
produced partons.

As an example of the dependence on the collision energy,
$\sqrt{s_{NN}}$, Figure~\ref{fig:chargedLHC} shows an extension of the
calculation to the CERN LHC, with $\sqrt{s_{NN}}=5500$ GeV for the
heavy ion collisions.  In the calculation of initial conditions from
the primary parton interactions, all parameters are fixed except for
the collision energy.  The saturation scale changes from $p_{\rm sat}=1.16$
GeV at RHIC energy to $p_{\rm sat}=2.03$ GeV at the LHC. The total
multiplicity increases from $dN{\rm tot}/dy \approx 1000$ at RHIC to
4500 at the LHC. The initial thermal densities are higher and lead to
longer expansion and stronger transverse flow at the decoupling. This
is seen in the change of the region where the component of thermal
particles goes over to the component of particles from jet
fragmentation with energy loss. At RHIC the transition is centered
around $p_T\sim3$ GeV, whereas at the LHC it is predicted to be
$p_T\sim5$ GeV. The larger $p_T$ region where thermal particles
dominate should also be seen in elliptic flow. At the present RHIC
energies, the hydrodynamic predictions of elliptic flow start to
overshoot the data above $p_T\sim2$ GeV, but if the thermal component
grows as predicted in Figure~\ref{fig:chargedLHC}, the elliptic flow
parameter $v_2$ should follow the hydrodynamical calculation to larger
transverse momenta.

%%%%%%%%%%%%%%%
%%%%%%%%%%%%%%%%%%%%%%%%%%%%%%%%%%%%%%%%%%%%%%%%%%%%%%%%%%
   \subsection{Elliptic Flow}
     \label{elliptic}
%%%%%%%%%%%%%%%%%%%%%%%%%%%%%%%%%%%%%%%%%%%%%%%%%%%%%%%%%%

The particle production in primary collisions is azimuthally
isotropic, whereas the reaction zone in noncentral collisions is not,
but has an elongated shape. If produced particles rescatter, the
particles moving in the direction of the longer axis of the reaction
zone are more likely to change their direction than the particles
moving in the direction of the shorter axis. Therefore the observed
emission pattern of particles will be azimuthally anisotropic, and the
more frequent the rescattering, the more anisotropic the particle
distribution.

In this way, the anisotropy of the final particle distribution is a
measurement of the frequency of rescatterings during the dense phase
of the collision. This anisotropy can be quantified as the
coefficients of the Fourier expansion of the azimuthal particle
distribution~\cite{Voloshin:1994}:
\bea
 \label{fourier}
  \frac{\mathrm{d}N}{\mathrm{d}y\mathrm{d}\phi_p}
    & = & \frac{\mathrm{d}N}{2\pi \mathrm{d}y}(1+2v_1\cos (\phi-\phi_R)
                              +2v_2\cos 2(\phi-\phi_R) + \cdots), \nonumber \\
  \frac{\mathrm{d}N}{\mathrm{d}y \mathrm{d}p_T \mathrm{d}\phi_p}
   & = & \frac{\mathrm{d}N}{2\pi \mathrm{d}y \mathrm{d}p_T}
                            (1+2v_1(p_T)\cos (\phi-\phi_R) \nonumber \\
   &   & \qquad\qquad\quad +{}2v_2(p_T)\cos 2(\phi-\phi_R) + \cdots),
\eea
where $\phi_R$ is the azimuthal angle of the event plane (the plane
spanned by the beam direction and the impact parameter). Assuming that
the experimental uncertainties in event-plane reconstruction can be
corrected for, each event can be rotated such that $\phi_R=0$.  The
first and second coefficient of the expansion, $v_1$ and $v_2$, are
usually referred to as directed and elliptic flow, respectively.
Because the system is usually thinner in the direction parallel to the
impact parameter, the in-plane direction, than in the out-of-plane
direction, the value of $v_2$ is positive.

At midrapidity, all uneven coefficients are zero owing to symmetry. At
SPS and RHIC energies, the directed flow, $v_1$, is expected to be
very small and most of the experimental and theoretical interest has
been directed toward measuring and analyzing the elliptic flow, $v_2$.
Recently the higher harmonics, $v_4, v_6$ and $v_8$, have also been
measured~\cite{Kolb:2003,Star-v4,Art-QM04}.

In a hydrodynamic picture, the buildup of momentum anisotropies is
easy to understand in terms of pressure gradients. The average
pressure gradient between the center of the system and the surrounding
vacuum is larger in the direction where the collision system is
thinner. Therefore, the collective flow is stronger in that direction
and more particles are emitted there than in the orthogonal direction,
where the collision system is longer.

As mentioned above, the more the particles rescatter, the larger the
observed anisotropy. Because hydrodynamics assumes practically
infinite scattering rate and zero mean free path, it is often assumed
to give an upper limit of anisotropy at fixed impact
parameter~\cite{Heinz}. However, this upper limit depends on the EoS
and, in principle, it is possible that hydrodynamical description with
very soft EoS would give a smaller anisotropy than, for example, a
microscopic cascade description~\cite{Molnar:2004yh}.

If the freeze-out happens at the same temperature for all particle
species, a signature of hydrodynamic flow is that the heavier the
particle, the flatter the slope of its $p_T$-spectrum. Similarly, the
$p_T$-averaged elliptic flow $v_2$ increases when particle mass
increases. However, the $p_T$-differential elliptic flow $v_2(p_T)$
has the opposite behavior: The heavier the particle, the smaller the
anisotropy at fixed $p_T$. The apparent discrepancy has a simple
explanation: $v_2$ is not an additive quantity, but when
$p_T$-averaged $v_2$ is calculated from $p_T$-differential $v_2(p_T)$,
the latter is weighted by the particle distribution. Thus, the flatter
$p_T$ distribution of a heavier particle weights more the high-$p_T$
region, where $v_2(p_T)$ is larger. Therefore, even if $v_2(p_T)$ is
smaller at all $p_T$ for a heavier particle, the $p_T$-averaged $v_2$
can be larger than $v_2$ for a light particle. Whether this happens in
practice and how large the differences are depend on the details of
the flow profile, i.e.\ expansion dynamics, and the resonance decays.

The mass ordering of the low-$p_T$ anisotropy has its origin in the
behavior of boosted particle distributions. As is well-known,
transverse flow shifts the $p_T$-dis\-tri\-bu\-tions to larger values
of $p_T$. In the extreme case in which the speed of the collective
motion is the same everywhere, as in the case of a thin shell
expanding with a velocity $v$, the particle distribution develops a
maximum at some finite $p_T$ (the so-called blast wave
peak~\cite{Siemens}) and a local minimum at $p_T=0$. The heavier the
particle, the larger the $p_T$ where the distribution peaks. Compared
with a case without transverse flow, the particle yield is thus
depleted at low $p_T$. The heavier the particle and the larger the
flow velocity $v_T$, the larger the depletion. Correspondingly, if the
flow velocity is larger in the in-plane than in the out-of-plane
direction, the low-$p_T$ depletion is larger for particles moving in
the in-plane direction than the out-of-plane direction and the overall
anisotropy, $v_2$, is reduced. This reduction and the range in which
it occurs increases with the particle mass and average transverse flow
velocity. In the extreme case of a thin expanding shell, this
reduction can be so strong that it reverses the sign of the anisotropy
and $v_2$ becomes negative. When the thin shell is replaced by a more
realistic velocity profile, the peak in transverse-momentum
distribution disappears. Similarly, a more realistic velocity
distribution weakens the reduction of $v_2$ at low $p_T$, but the mass
ordering of $v_2$ at low $p_T$ remains. Whether some particles depict
positive or negative $v_2$ at low $p_T$ depends on the details of the
flow velocity of the source.

For relativistic $p_T > m$, the particle mass does not play any role
in the thermal distribution and, consequently, $v_2(p_T)$ of different
particles converge. In a simple model in which the flow-velocity
profile is replaced by its average value, $v_2(p_T)$ increases with
$p_T$ and approaches unity asymptotically. The details of the
flow-velocity profile can change this behavior, but so far no
hydrodynamical calculation has reproduced the experimentally observed
saturation of elliptic flow.

 \subsubsection{CENTRALITY DEPENDENCE}

\begin{figure}
   \hfill
 \begin{minipage}[t]{65mm}
    \epsfysize 55mm \epsfbox{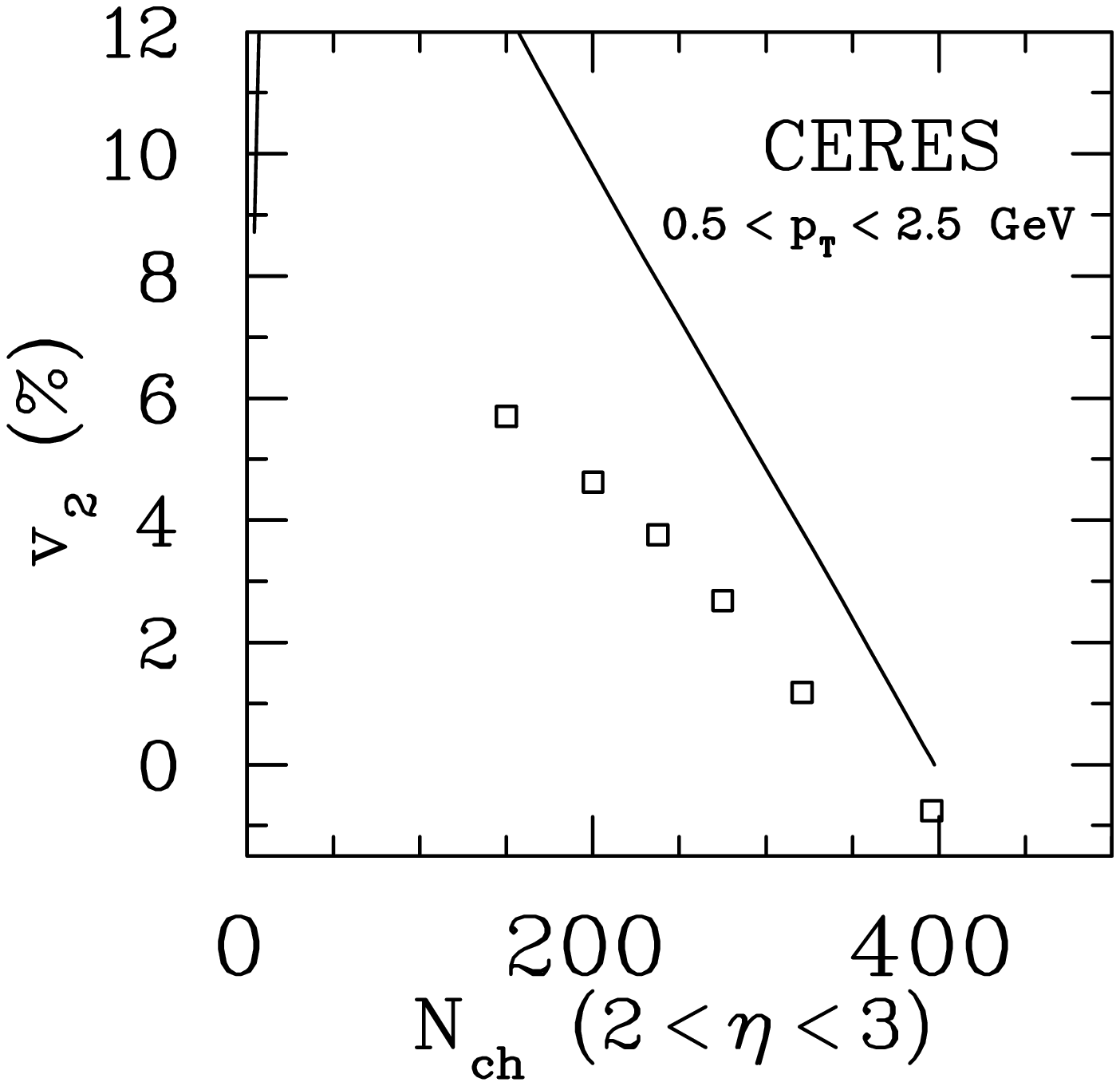}
 \end{minipage}
   \hfill
 \begin{minipage}[t]{65mm}
    \epsfysize 55mm \epsfbox{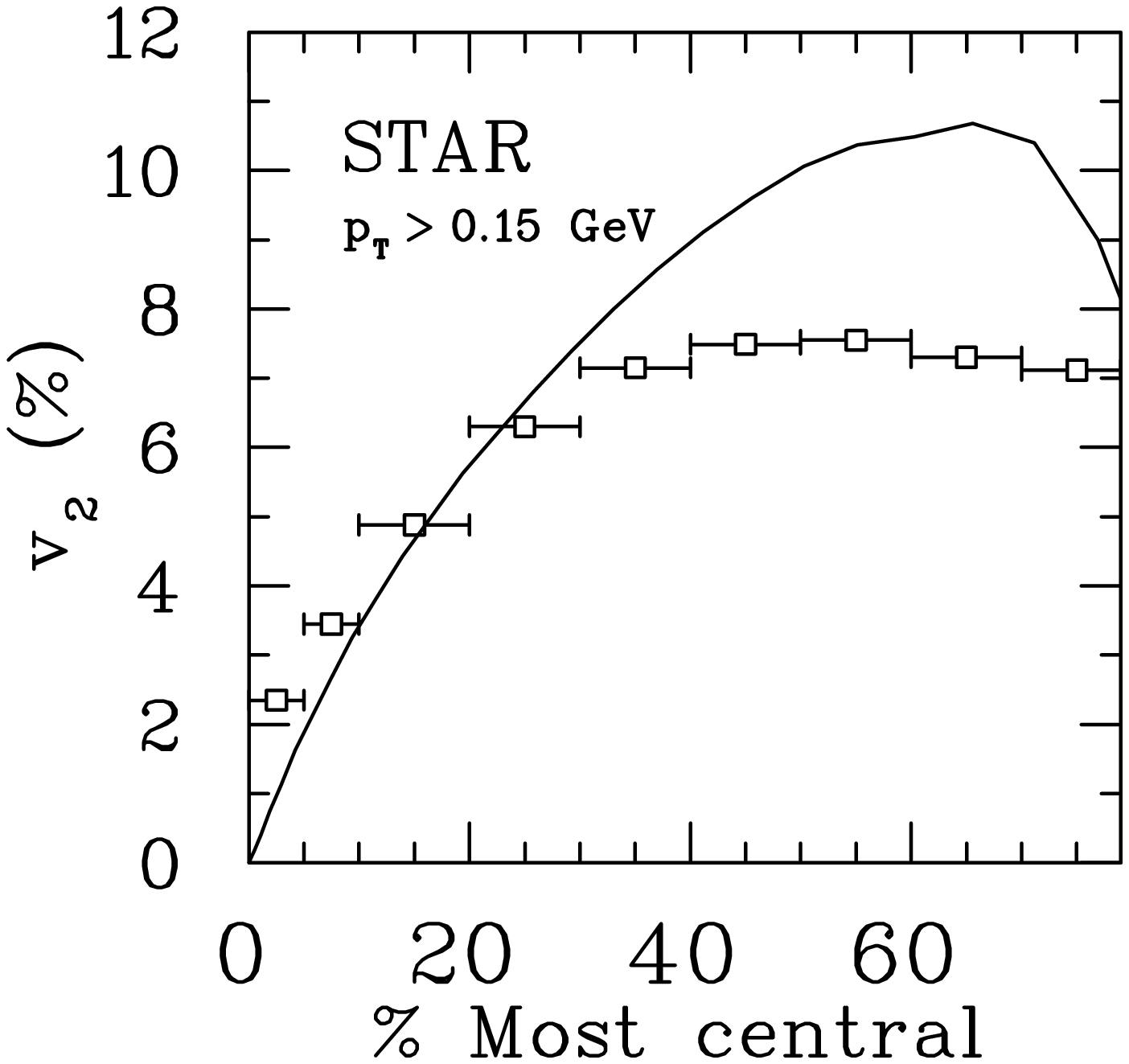}
 \end{minipage}
   \hfill
 \caption{\protect\small
          Elliptic flow $v_2$ of charged hadrons as a function of
          centrality at (left panel) Pb+Au collisions at $\ssNN=17.3$
          GeV measured by CERES~\cite{ceres-v2} and at (right panel)
          Au+Au collisions at $\ssNN=200$ GeV measured by
          STAR~\cite{Star-v4}. Hydrodynamical calculations are taken
          from References~\cite{ceres-v2,Huovinen-EoS}. Note that the
          hydrodynamical results are not directly comparable because
          of different $p_T$ cuts implemented.}
 \label{v2cent}
\end{figure}
In a hydrodynamic description, the final anisotropy of particles is
almost directly proportional to the geometrical anisotropy of the
initial state~\cite{Peterinpitka}. The proportionality is, however,
nontrivial and depends on the applied decoupling criterion. When the
impact parameter increases and the collision becomes more peripheral,
the collision system becomes more and more elongated and its geometric
anisotropy increases. We can thus expect the observed momentum
anisotropy to increase as well. The data in Figure~\ref{v2cent} shows
increasing elliptic flow with decreasing centrality both at SPS and
RHIC energies, but the magnitude of the flow differs from the
hydrodynamical result. At SPS energy the data is consistently below
the calculation, whereas at RHIC energy ($\sqrt{s_{NN}} = 200$ GeV in
Figure~\ref{v2cent}) the data is reproduced up to semi-central
collisions but is below the calculation at peripheral collisions.

The failure of hydrodynamics to describe the anisotropy in most
peripheral collisions and at SPS energy is often explained by a lack
of necessary thermalization owing to the small size and/or particle
number of the collision system~\cite{Heinz-eta}. An alternative
explanation assumes that the initial partonic state is sufficiently
thermalized, but the final hadronic state has such a large viscosity
that it cannot be modeled using ideal-fluid dynamics~\cite{Teaney-RQMD}.
The latter approach has been tested using so-called hybrid models in
which the plasma phase and phase transition are described using ideal
hydrodynamics, but the hadron phase is described using a cascade
model~\cite{Bass:2000ib,Teaney-RQMD,Nonaka:2005aj,Hirano:2005xf}. The
centrality dependence of $p_T$-averaged elliptic flow at RHIC has been
reproduced nicely using such a hybrid
approach~\cite{Hirano:2005xf,Teaney-RQMD}, but the results for $p_T$
differential $v_2$ at $\ssNN = 200$ GeV collision energy are not
available at the time of this writing.

At the beginning of this section it was argued that hydrodynamics
leads to the largest possible anisotropy. However, at most central
collisions at RHIC, the data tends to be \emph{above} the
hydrodynamical calculation. Fluctuations in the shape of the initial
system may explain this. Owing to these fluctuations, the initial
shape of some events in almost-central collisions can be in-plane
elongated, even if the shape on most events is out-of-plane elongated.
Thus elliptic flow is negative in some events, but because
experimental analysis measures the magnitude of the anisotropy, not
its sign, elliptic flow is measured as positive in all events and the
measured value is larger than the average value. The initial state of
hydrodynamical calculation, however, is an average initial state in
which fluctuations of the spatial anisotropy cancel each other and the
calculated anisotropy is smaller than measured~\cite{Miller}.
Preliminary calculations in which the initial-state fluctuations are
included favor this interpretation by leading to better reproduction
of the data~\cite{Hama}.

The general trend is that a stiffer EoS and a lower freeze-out
temperature lead to larger $p_T$-averaged flow if nothing else in the
model is changed. This also changes the single particle distributions.
If these are still required to fit the data, additional changes are
required. For example, a stiffer EoS usually necessitates a higher
freeze-out temperature. The combined effect largely cancels and the
final $p_T$-averaged anisotropy is almost unchanged in semi-peripheral
collisions in which a hydrodynamical description works
best~\cite{Huovinen-EoS}.

 \subsubsection{MOMENTUM and PARTICLE SPECIES DEPENDENCE}

Hydrodynamic calculations at SPS and RHIC energies lead to anisotropy,
which increases with increasing $p_T$ and approaches unity
asymptotically. Simple parame\-tri\-zations of flow (so-called
blast-wave models) also lead to this kind of behavior, which differs
from experimental observations in which $v_2$ saturates at high
$p_T$. At mid rapidity, the agreement between the data and
calculations depends on energy in the same way as for the centrality
dependence: At SPS energies the calculations overestimate the data,
whereas at RHIC energy a good agreement can be reached. The $p_T$
range in which the data can be reproduced depends on the particle
species. Charged hadrons and pions can be fit up to $p_T \approx 1.5$
GeV, whereas protons follow the calculations up to $p_T = 2.5$--3 GeV
in minimum-bias collisions. This behavior is qualitatively similar to
that seen in Figures~\ref{fig:positive130} and~\ref{fig:positive200}
for transverse momentum spectra, where hydrodynamically calculated
spectra fit the data up to $p_T\sim 3$ GeV.

A hydrodynamic description predicts a characteristic mass dependence
of elliptic flow at low $p_T$. The higher the mass, the lower the
$v_2$. How large this difference is depends on the details of the flow
profile and therefore on the EoS. If chemical equilibrium in the
hadronic phase is assumed, the differential anisotropy of pions can be
well reproduced when the $p_T$ spectra of pions is reproduced. In such
a case, the proton $v_2(p_T)$ depicts sensitivity to the phase
transition. If the phase transition takes place in a narrow
temperature interval and has large latent heat, the proton
differential anisotropy can almost be reproduced. If there is no phase
transition, the calculated proton anisotropy is clearly above the
data~\cite{Huovinen-EoS}. The status of strange particles is less
satisfactory. Kaons and $\Lambda$-baryons show similar dependence on
the EoS than protons, but the difference between the data and
hydrodynamical calculation is larger than in the case of
protons~\cite{Huovinen-EoS}.

\begin{figure}
   \hfill
 \begin{minipage}[t]{65mm}
    \epsfysize 55mm \epsfbox{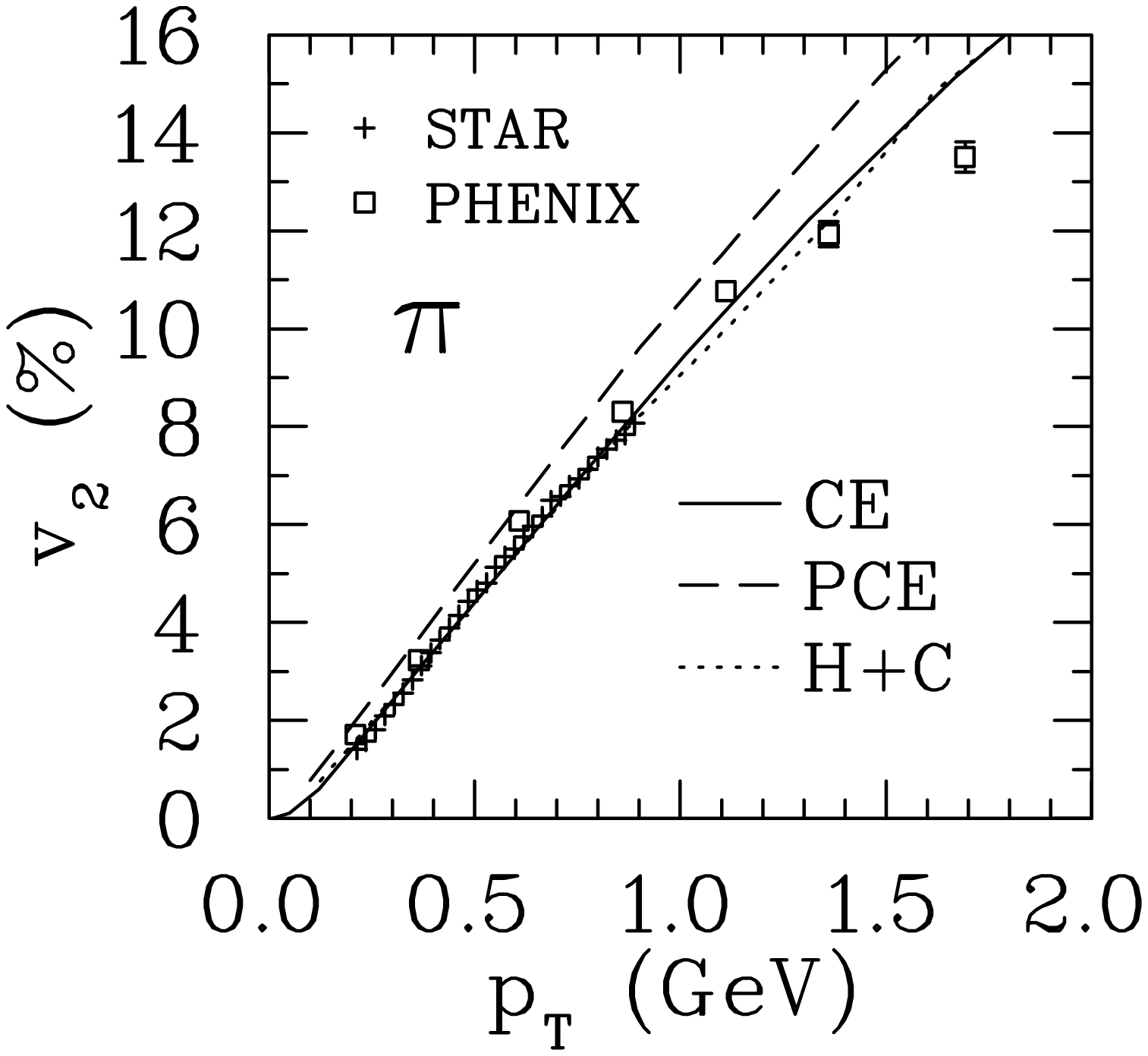}
 \end{minipage}
   \hfill
 \begin{minipage}[t]{65mm}
    \epsfysize 55mm \epsfbox{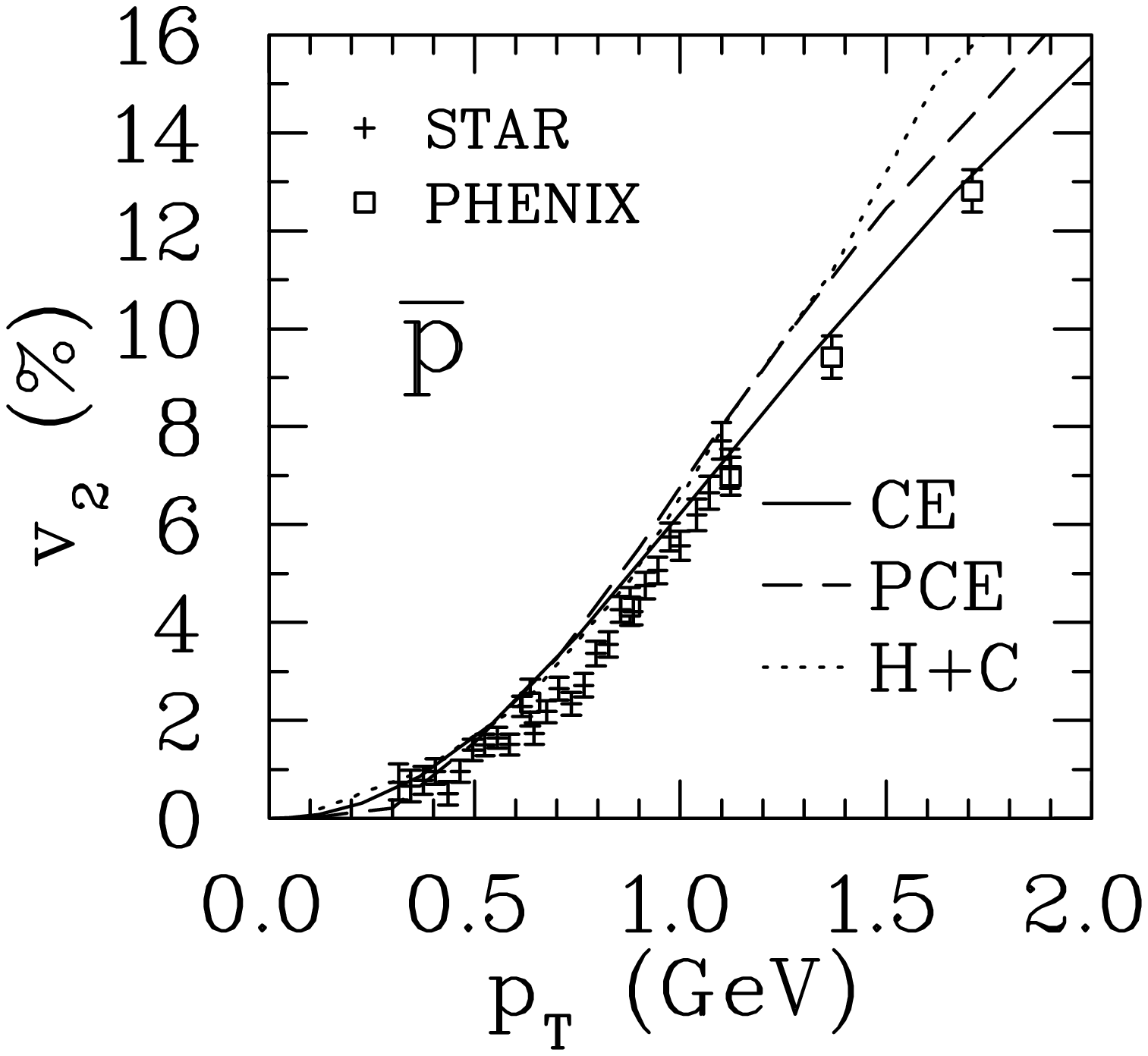}
 \end{minipage}
   \hfill
 \caption{\protect \small
          Pion (left panel) and antiproton (right panel) $v_2(p_T)$ at
          midrapidity in minimum-bias Au+Au collisions at $\ssNN=200$
          GeV.  Hydrodynamical results are labeled CE for
          chemical-equilibrium result~\cite{Huovinen-EoS}, PCE for
          chemical-nonequilibrium result~(\cite{Hirano2,Hirano:2005xf};
          T.~Hirano, personal communication) and H+C for hydro+cascade
          hybrid model at $\ssNN=130$ GeV energy~\cite{Teaney-RQMD}.
          The data are taken by STAR~\cite{Star-v2-200} and
          PHENIX~\cite{Adler:2003kt}.}
 \label{v2pt}
\end{figure}

Unfortunately, it is not yet possible to use the apparent sensitivity
of proton $v_2(p_T)$ to the EoS to quantitatively constrain the EoS.
If the requirement of chemical equilibrium is relaxed and one uses
separate chemical and kinetic freeze-out temperatures, the fit to pion
$v_2(p_T)$ is lost (see Figure~\ref{v2pt} and Reference~\cite{Hirano2}).
However, if the hadronic phase is described using the RQMD transport
model, as in Reference~\cite{Teaney-RQMD} for the $\ssNN=130$ GeV
collision energy, the yields are correct and the $v_2(p_T)$ is
described as well as in the case of chemical equilibrium. Thus, there
is a considerable uncertainty in the description of the hadronic stage
of the evolution, which makes it impossible to draw final conclusions
about the EoS needed to describe the differential anisotropy.

Another uncertainty here is the effect of viscosity. The ability of
ideal-fluid hydrodynamics to reproduce the $v_2$ data at RHIC has been
interpreted to mean that the value of shear viscosity in QGP is
particularly low~\cite{Hirano:2005wx,Gyulassy:2004zy}.  However,
viscosity has been estimated to decrease elliptic
flow~\cite{Teaney-visco}, and chemical non-equilibrium increases
elliptic flow~\cite{Hirano2,Peter-Ralf}. Hirano \& Gyulassy have
argued that the plasma has sufficiently low viscosity to allow its
modeling using ideal hydrodynamics, but the dissipative effects in the
hadronic phase are the reason for the failure of chemically
non-equilibrium hydrodynamics to describe the
data~\cite{Hirano:2005wx}. This claim seems to be validated by the
ability of hybrid models to describe the data, but final conclusions
must wait for the complete results of hybrid calculations and the
results of viscous calculations~\cite{Muronga,Teaney:2004qa}.

 \subsubsection{PSEUDORAPIDITY DEPENDENCE}

The $p_T$-averaged elliptic flow has quite a strong dependence on
pseudorapidity~\cite{Phobos130,Phobos200,Star-v2-200,Staszel:2005aw}. In
a narrow region around midrapidity, $|\eta| < 1$, elliptic flow
remains approximately constant~\cite{Star-4cor,Star-v2-200} but
decreases strongly towards larger pseudorapidities. However, charged
particle multiplicity depicts much wider plateau around midrapidity
than elliptic flow~\cite{Phobos-multi}.

The purely hydrodynamical calculations have not reproduced the
pseudorapidity dependence of elliptic flow, although the same
calculations reproduce the multiplicity as a function of
pseudorapidity~\cite{Hirano1,Hirano2}. However, a hybrid model that
reproduces the centrality dependence of elliptic flow also gives a
reasonable description of its pseudorapidity
dependence~\cite{Hirano:2005xf}. The failure of ideal-fluid
hydrodynamics has been interpreted in the same way as in the case of
centrality dependence -- either as an incomplete thermalization of the
system at large rapidities from the beginning~\cite{Heinz-eta} or as
an effect of viscosity and incomplete thermalization at the late
stage~\cite{Hirano:2005xf}. There are also other open questions in the
hydrodynamic treatment that may affect the results: The initial shape
of the system, large deviations from boost-invariant flow, and
different thermalization time and freeze-out temperature at different
rapidities could all affect the final anisotropy and are mostly
unexplored. It is therefore possible that the thermalized region where
hydrodynamics works at RHIC energy is relatively narrow in rapidity,
but final conclusions cannot be drawn yet.

%%%%%%%%
%%%%%%%%%%%%%%%%%%%%%%%%%%%%%%%%%%%%%%%%%%%%%%%%%%%%%%%%%%%%%%%%%%%%
  \subsection{Two-Particle Bose-Einstein Correlations}

%%%%%%%%%%%%%%%%%%%%%%%%%%%%%%%%%%%%%%%%%%%%%%%%%%%%%%%%%%%%%%%%%%%%

Information about the space-time structure of the system formed in a
heavy-ion collision can be obtained by measuring the low-momentum
correlations of identical particles. For bosons these correlations are
called Bose-Einstein correlations and the method for their
interpretation is called HBT interferometry according to the
originators of this method~\cite{twiss}. Here we show only the basics
of the HBT formalism as applied to hydrodynamical models and its most
important results. A detailed explanation about this technique can be
found in Reference~\cite{hbt-report}, and the present status is
discussed in recent reviews~\cite{Lisa,Tomasik}. HBT in hydrodynamical
context is more throughly discussed in Reference~\cite{Peter-review}.

Intensity interferometry is based on an analysis of the two-particle
momentum correlation function,
\be
  C(\mbox{\boldmath $q$},\mbox{\boldmath $K$})
     = \frac{E_1 E_2\frac{\mathrm{d}N}{\mathrm{d}^3\pb_1\mathrm{d}^3\pb_2}}
            {E_1\frac{\mathrm{d}N}{\mathrm{d}\pb_1}
             E_2\frac{\mathrm{d}N}{\mathrm{d}\pb_2}},
\ee
that is, the ratio of a two-particle distribution and a product of two
one-particle distributions. The correlator is usually written in terms
of the momentum difference between the two particles, $q = p_1 - p_2$,
and their average momentum, $K = \frac{1}{2}(p_1+p_2)$. If the
particles are emitted independently (``chaotic source'') and propagate
freely from the source to the detector, the two-particle distribution
is not equal to the product of one-particle distributions. At small
values of relative momentum $q$, it is larger than the product of
one-particle distributions owing to quantum statistical (wave-function
symmetrization) effects.

If there are no final-state interactions (or the spectra are corrected
for them), the two-particle correlator
$C(\mbox{\boldmath $q$},\mbox{\boldmath $K$})$ is related to the
emission function $S(x,K)$:
\be
  C(\mbox{\boldmath $q$},\mbox{\boldmath $K$}) \approx
   1 + \left|\frac{\int \mathrm{d}^4x\, S(x,K) e^{\mathrm{i} q\cdot x}}
                  {\int \mathrm{d}^4x\, S(x,K)}\right|^2.
 \label{CwithS}
\ee

The emission function $S(x,K)$ is the Wigner phase-space density of
the emitting source. In the derivation of Equation \ref{CwithS} the
emission function is assumed to be sufficiently smooth, i.e.\
$S(x,K) \approx S(x,K+\frac{1}{2}q)$~(see Reference~\cite{hbt-report}).
Because both $p_1$ and $p_2$ are on-shell, the average momentum $K$
is, strictly speaking, off-shell. In practice, however, on-shell
approximation for $K$ is used:
$K_0 \approx \sqrt{\mbox{\boldmath $K$}^2 + m^2}$.

In the hydrodynamic approach the quantum-mechanical Wigner phase-space
density is replaced by a classical phase-space density at the time of
freeze-out. When Cooper-Frye formalism is applied, it is given by
\be
 S(x,K) = \frac{g}{(2\pi)^3}
          \int\frac{\mathrm{d}\sigma_{\mu}(x') K^\mu \delta^4(x-x')}
                   {\exp\{[K\cdot u(x')-\mu(x')]/T(x')\}\pm1}.
\ee

It is not possible to define uniquely the source function $S(x,K)$
from the measured correlation function
$C(\mbox{\boldmath $q$},\mbox{\boldmath $K$})$. The experimental data
of two-particle correlations are therefore presented using some ansatz
for the the source function. Usually this is done using a Gaussian
form for the correlator. If the collision system is further
approximated to be boost-invariant, the correlator for central
collisions can be written in a particularly simple form in terms of
three HBT radii:
\be
 C(\mbox{\boldmath $q$},\mbox{\boldmath $K$}) \approx
    1+ \exp[-R_o^2(K_T)q_o^2 - R_s^2(K_T)q_s^2 - R_l^2(K_T)q_l^2].
\ee
In this so-called Bertsch-Pratt parametrization, the coordinate
directions are defined in such a way that out- ($R_o$) and
long-direction ($R_l$) are parallel to $\mbox{\boldmath $K$}_T$ and
beam, respectively, whereas the side-direction ($R_s$) is
perpendicular to both $\mbox{\boldmath $K$}_T$ and beam. In a
boost-invariant approximation, the radii depend only on the magnitude
$K_T$ because the particle emission in central collisions does not
depend on the azimuthal angle $\phi$ and boost-invariance means that
there cannot be any rapidity dependence.

These radii do not correspond to the actual physical size of the
source, but rather characterize so-called regions of homogeneity, the
regions where particles with particular $p_T$ are most likely
emitted. For a Gaussian source, the HBT radii measure the following
different combinations of space-time \emph{variances} of the
system~\cite{Peter-review}:
\bea
  R_s^2(K_T) & = & \langle\tilde{x}_s^2\rangle(K_T) \label{xside} \\
  R_o^2(K_T) & = & \langle(\tilde{x}_o - \beta_\perp\tilde{t})^2\rangle(K_T)
                      \label{xout} \\
  R_l^2(K_T) & = & \langle\tilde{x}_l^2\rangle(K_T), \label{xlong}
\eea
where $\beta_\perp = K_T/K^0$ is the transverse pair velocity, and
space-time coordinates $\tilde{x}$ are defined as distances from the
``effective source center'' $\tilde{x}^\mu(K_T) = x^\mu - \langle
x^\mu\rangle(K_T)$, where brackets denote weighted averages over the
source function $S(x,K)$:
\be
 \langle f(x)\rangle (K) = \frac{\int d^4x\, f(x) S(x,K)}{\int d^4x\, S(x,K)}.
\ee
The radii are thus independent of the actual coordinates of the
emission, but are sensitive to variances of the geometry.

Hydrodynamic calculations for RHIC energies predicted that a phase
transition from a plasma to a hadron gas would increase the lifetime
of the system~\cite{Rischke-hbt}. The long lifetime would increase the
$\beta_\perp\tilde{t}$ term in Equation \ref{xout} and thus increase
the $R_o$-radius and make the ratio $R_o/R_s$ large. The experimental
data, however, shows no sign of this kind of effect and yields a ratio
of $R_o/R_s \approx 1$.

\begin{figure}
 \epsfxsize11cm \centerline{\epsfbox{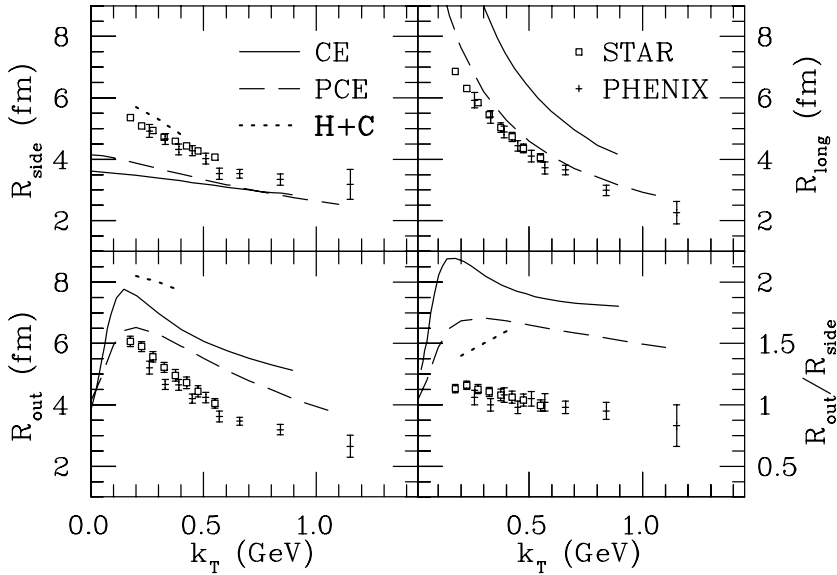}}
 \caption{\protect \small
          Pion source radii in Au+Au collisions at $\ssNN=200$
          GeV. Hydrodynamical results are labeled CE for
          boost-invariant chemical-equilibrium result~\cite{Heinz2},
          PCE for nonboost-invariant result with separate chemical and
          kinetic freeze-outs~(\cite{Hirano2,Hirano:2005xf};
          T.\ Hirano, personal communication) and H+C for hydro+cascade
          hybrid model~\cite{Soff}. The data are taken by the
          STAR~\cite{Adams:2004yc} and PHENIX~\cite{Adler:2004rq}
          collaborations.}
 \label{hbtkuva}
\end{figure}

Figure~\ref{hbtkuva} shows some of the hydrodynamic calculations for
the HBT radii at RHIC energy.
Hydrodynamic calculations with assumed boost-invariance, chemical
equilibrium, and a first-order phase transition (CE, solid line) tend
to lead to a too small sideward radius $R_{s}$, too large outward and
longitudinal radii $R_o$ and $R_l$, respectively, and especially to a
too large ratio $R_o/R_s$~\cite{Heinz2}. Both $R_o$ and $R_l$ can be
made smaller if the system decouples sooner, i.e.\ in higher
temperature, but such an approach leaves $R_s$ basically unchanged and
distorts the single particle spectra~\cite{Heinz2,Peter-review}. To a
lesser extent, the same effect can be achieved by decreasing the
initial time or increasing the transverse flow by a non-zero initial
velocity field, but neither of these approaches changes the too small
$R_s$~\cite{Heinz2}.

Another way to reduce the longitudinal radius $R_l$ is to relax the
boost-invariant approximation~\cite{Hirano2,Morita}. When this
approach is used with an EoS with separate chemical and kinetic
freeze-outs (PCE, dashed line), $R_l$ is close to the data. Both $R_o$
and $R_s$ move closer to the data but are still too large and small,
respectively. This approach also leads to problems with the elliptic
anisotropy (see chapter~\ref{elliptic}).

An approach that brings $R_s$ close to the data chooses a wide but
flat initial distribution~\cite{Morita}, which leads to slower buildup
of flow from an initially larger source. In that case, the problem is
again $R_o$, which is too large. This is expected because $R_o$ is
sensitive to the lifetime of the system, which becomes relatively
large in this approach.

One way to reduce the lifetime of the system and thus $R_o$ is to
change the EoS. As mentioned, one of the suggested signatures of a
first-order phase transition is a long lifetime and large $R_o$. If
one uses an EoS with a smooth crossover instead of a first-order phase
transition, $R_o$ decreases and $R_s$ increases~\cite{Zschiesche}.
Unfortunately, even in that case, no good fit to the data is
achieved. It is also questionable how this change in EoS would affect
the elliptic anisotropy~\cite{Huovinen-EoS}.

Grassi et al.\ have suggested that the discrepancy between the data
and calculations is due to too simple a treatment of freeze-out on a
sharp hypersurface, and a more realistic continuous emission of
particles would lead to better results~\cite{Grassi}. However, when
this is accounted for effectively in hybrid models in which the
hadronic stage is described using a cascade transport model, the
results are even worse~\cite{Soff}. The particles are emitted from
larger, longer-lasting volume than in a simple hydrodynamic
description, and correspondingly, $R_s$ is larger and reproduces the
data (H+C, dotted line in Figure~\ref{hbtkuva}). Unfortunately, the
longer lifetime also leads to an even larger $R_o$.

Another possible reason for the discrepancy between the data and
calculations is viscosity~\cite{Dumitru,Teaney-visco}. Initial
calculations~\cite{Muronga} show that it has the desired effects, but
whether they are large enough remains to be seen. Again, the effect of
viscosity on elliptic flow is large, and it is unknown if a viscous
model could reproduce both the HBT radii and elliptic anisotropy.

%%%%%%%%%%%%%%%%%%%%%%%%
%%%%%%%%%%%%%%%%%%%%%%%%%%%%%%%%%%%%%%%%%%%%%%%%%%%%%%%%%%%%%%%%%%%%%
\section{ELECTROMAGNETIC EMISSION}
  \label{em}
%%%%%%%%%%%%%%%%%%%%%%%%%%%%%%%%%%%%%%%%%%%%%%%%%%%%%%%%%%%%%%%%%%%%%
%%%%%%%%%%%%%%%%%%%%%%%

All the observables described in the previous section \ref{hadrons}
are hadronic observables. By definition, the hadrons of the system
interact with each other, and the distributions and yields of hadrons
are fixed late in the evolution of the system, when interactions
cease, the distributions and yields freeze out, and the particles
decouple.  Therefore, those observables characterize the properties of
the particle-emitting source at the end of the system evolution, but
not the history of the system during the evolution. In principle, it
is possible to have very different dynamics producing similar final
states.

Possible observables that are sensitive to the entire evolution of the
collision system are photon and lepton-pair distributions. Because
these particles interact only electromagnetically, their mean free
paths are much longer than those of hadrons. They can thus escape the
collision system without rescattering and carry information about the
conditions in which they were formed. However, the photon and dilepton
spectra get contributions from all stages of the evolution, which
makes it difficult to disentangle the signal coming from the hot,
dense stage of the collision. To describe the different contributions
to electromagnetic spectra, we follow the terminology of
Reference~\cite{yellow}:
\begin{itemize}
 \item prompt photons and leptons are produced in the primary
       collisions of incoming partons;
 \item thermal photons and leptons are emitted in the collisions of
       quarks and gluons during the plasma phase and in the collisions
       of hadrons in the hadronic phase;
 \item decay photons and leptons are decay products of hadrons;
 \item direct photons and leptons are the sum of prompt and thermal photons.
\end{itemize}
The thermal photon production depends strongly on temperature via the
factor $\exp(-p_T/T)$. Therefore the early stage, when the matter is
hottest, should dominate the photon emission, and the measurement of
photon spectra should be an effective thermometer for the temperature
achieved in the collision. However, prompt photons follow a power-law
distribution $p_T^{-n}$ and dominate at high $p_T$.

The hydrodynamic model can be used to calculate the thermal and decay
contributions, but the prompt photons and leptons require a separate
pQCD calculation. The calculation of decay photons is relatively
straightforward, and it proceeds in the same way as the calculation of
hadron spectra from resonance decays described in
section~\ref{sec:spectra_hadron}. One calculates the distribution of
hadrons at freeze-out and applies the relevant decay kinematics and
branching ratios to get the spectra of decay photons and leptons.  To
calculate the thermal yield, the production rate of photons or leptons
in a thermal system, $\mathrm{d}R/\mathrm{d}^3p(E,T,\mu)$, has to be
integrated over the space-time volume of the system:
\bea
  E\frac{dN}{d^3p} = \int d^4 x \, \Big\{ && \;\;\;
  w(\varepsilon,\rho_{\rm B}) \; E\frac{dR^{\rm QGP}}{d^3p}(p\cdot u, T,
  \mu_{\rm B}) \nonumber\\ && \!\!\!\!\!\!\!\!\!\!  +\left[1 -
  w(\varepsilon,\rho_{\rm B})\right] E\frac{dR^{\rm HG}}{d^3p}(p\cdot u,
  T,\mu_{\rm B}) \Big\} \: ,
 \label{stint}
\eea
where the production in a plasma and in a hadron gas (HG) is written
separately.  The factor $w(\varepsilon,\rho_{\rm B})$, which expresses
the volume fraction of plasma, is unity in the plasma phase, zero in
HG and between unity and zero in a mixed phase. Hydrodynamics provides
the space-time evolution of the system, whereas the production rates
in thermal matter are an input independent of hydrodynamics.

In the plasma phase, the photon production is dominated by the QCD
Compton and annihilation reactions, $qg \to q\gamma$, $\bar{q}g \to
\bar{q}\gamma$, and $q\bar{q} \to g\gamma$. In lowest order, the
production rate due to these processes was calculated in
References~\cite{Kapusta,Baier}. However, some formally higher-order
processes are strongly enhanced by collinear singularities and also
contribute to order $\alpha_{s}$ \cite{Aurenche,Steffen}. The
resummation of these contributions was shown to be possible and was
carried out by Arnold et al.~\cite{Arnold:rate}, completing the order
$\alpha_s$ analysis of the photon-emission rate. A parametrization of
the rate was also given in Reference~\cite{Arnold:param}.

The calculation of the photon-production rate in a hot hadron gas is
less complete than the rate in a plasma owing to the multitude of
different hadron species and photon-producing interactions and owing
to the model dependence of the calculations (see
References~\cite{yellow,Gale-review}). The standard rate in the
literature is the one calculated in Reference~\cite{Kapusta}, where
photon production in scattering and decay processes
$\pi\pi\to\rho\gamma$, $\pi\rho\to\pi\gamma$, $\omega\to\pi\gamma$,
and $\rho\to\pi\pi\gamma$ was calculated using pseudoscalar-vector
Lagrangian with coupling constants determined from free $\rho$ and
$\omega$ decays.  These rates are often supplemented with a production
rate via $a_1$-mesons from Reference~\cite{a1}.

The role of different channels in photon production was further
studied using chiral Lagrangians~\cite{Song}. Unfortunately, it was
not possible to fix the model parameters unambiguously in this work,
which led to a factor of three uncertainty in the final rates. In the
context of dilepton production, it was later possible to fix the model
parameters much better~\cite{Gao}, and this approach was used in a
recent calculation by Turbide et al.~\cite{Turbide}. In that work, the
study was extended to cover photon emission from heavier meson
resonances, strange particles, and baryons. Another recent analysis of
photon production from a hadron gas was done by Haglin~\cite{Haglin},
who studied the effect of strange particles and higher-order processes
achieving a rate larger than the standard rate~\cite{Kapusta} by a
factor of two at large $q_T$ and by an order of magnitude at low $q_T$.

There are still uncertainties in the calculation of the
photon-production rate in hot hadron gas. Surprisingly, even after all
the improvements in the calculations, the statement made in
Reference~\cite{Kapusta} is still valid. At the same temperature, the
production rate per unit volume in a plasma and a hot hadron gas is
approximately equal, and they both ``shine as brightly''. However, the
emission rate per unit entropy is larger in hadron gas.

The main contribution to dilepton production in plasma comes from the
annihilation process $q\bar{q} \to l\bar{l}$. The rate calculated in
lowest order in a baryon-free plasma can be found in
textbooks~\cite{Wong} and was calculated for finite baryon chemical
potentials in Reference~\cite{Cleymans}. At small values of invariant
lepton-pair mass, corrections of the order $\alpha\alpha_s$ to this
rate become important~\cite{Pisarski}, but in heavy ion collisions
lepton pairs from Dalitz decays of final mesons produce a larger
background~\cite{Altherr}. Multi-loop calculations similar to those
done to calculate the photon rate have also been carried out for
high-$p_T$ pairs with small invariant mass~\cite{lepton-loop}. These
calculations have resulted in rates somewhat larger than the
first-order calculations.  First attempts to calculate lepton
production using a lattice-QCD formalism have also been
done~\cite{hilarate}. The preliminary results are quite close to the
perturbative rate, at least in some parts of the phase space.

The observation of large excess dileptons in the mass region below the
$\rho$-meson mass in Pb+Au collisions at $\sqrt{s_{NN}}=17.3$ GeV
energy at the CERN-SPS~\cite{ceres} has fueled considerable
theoretical interest in studying the lepton-pair emission in a hot
hadronic gas~\cite{Gale-review,Rapp-review}. The main problem of these
studies has been whether and how the properties of mesons change in
medium and how these changes are reflected in the rates. A rate
calculated by Gale \& Lichard~\cite{GaleLichard} using free-particle
properties is often used as a benchmark in comparisons with more
sophisticated approaches. In the calculations of Rapp et
al.~\cite{Rapp} and Eletsky et al.~\cite{Eletsky}, the basic
assumption is that the spectral density of $\rho$-meson changes in
medium. These calculations are technically very different, but produce
qualitatively similar rates~\cite{leptons01}. An alternative approach
pursued by Brown \& Rho~\cite{Brown-Rho} assumes that the $\rho$-meson
mass decreases in the medium.

\subsection{Photons at SPS}

Direct photon production in $\sqrt{s_{NN}} = 17.3$ GeV Pb+Pb
collisions at the CERN-SPS was measured by the WA98
collaboration~\cite{WA98photons}. Several authors have compared this
data with hydrodynamical
calculations~\cite{Srivastava,Alam,Peressounko,Chaudhuri,sps-photons}. All
authors agreed that the photon spectrum could be explained if one
assumes sufficiently hot ($T > 200$ MeV) initial state, but the
required initial temperature varied largely from $T\sim 200$
MeV~\cite{Alam} to $T=335$ MeV~\cite{Srivastava}. The large difference
is owing mainly to different assumptions in the calculations.

One factor that explains the largely varying initial temperature is
the use of different rates in a hadron gas. Alam et al.~\cite{Alam}
assumed in-medium modifications to hadron properties both in the EoS
and in production rates, which enhance the photon emission at lower
temperatures, allowing cooler initial state. The full order $\alpha_S$
rate for photon production in plasma was used only in the most recent
paper~\cite{sps-photons}, but at SPS energy the different rates in
plasma cause significant differences in the final yield only at
relatively large values of $p_T$.

The initial state of the system was also chosen in different ways in
different calculations. Especially, the assumption of finite
transverse flow velocity at the beginning of the hydrodynamic
evolution leads to lower temperatures. Because the rates are
proportional to $\exp(-p\cdot u/T)$, where $p$ is the four momentum of
the photon and $u$ is the flow four-velocity, stronger transverse flow
allows lower temperatures to produce equal yield at high $p_T$.
Peressounko \& Pokrovsky~\cite{Peressounko} argued the necessity of
such an initial flow, and Alam et al.~\cite{Alam} and
Chaudhuri~\cite{Chaudhuri} later studied its effects. It can be argued
that gradients in initial particle production would lead to buildup of
flow during thermalization, but it is very difficult to quantify how
large flow velocities could build up this way.  Peressounko \&
Pokrovsky also argued that the pion spectra especially necessitates
the initial flow, but the authors of this review have not been able to
fit the hadron spectra if initial transverse flow is assumed.

The hadron spectra were reproduced in
References~\cite{sps-photons,Peressounko}, and, when no initial
transverse flow is assumed, also in Reference~\cite{Alam}. In the
other two calculations~\cite{Srivastava,Chaudhuri}, the initial state
was only required to have the same entropy as the final-state
particles.  It is thus unknown whether these calculations are
consistent with the hadron data.

With the exception of Reference~\cite{sps-photons}, boost-invariant
hydrodynamics was used in these calculations. If high initial
temperature is required, this assumption leads to short initial, i.e.\
thermalization, time, $\tau \sim 0.2$ fm/$c$~\cite{Srivastava}. At SPS
energy it can be argued that such a short initial time is ambiguous
because the longitudinal extension of the colliding nuclei is larger
than $c\tau_0$.  This makes the application of boost-invariant
expansion uncertain for times $\tau<1$ fm/$c$. In
Reference~\cite{sps-photons} this problem was solved by using
nonboost-invariant hydrodynamics, where longitudinal geometry is
explicit and the initial time does not appear.  Also, the ambiguity in
choosing the initial state was studied and it was shown that several
EoSs and initial states reproduced both the hadron and photon
data~\cite{sps-photons,Huovinen98}.

\begin{figure}
   \hfill
 \begin{minipage}[t]{63mm}
   \epsfxsize 58mm \epsfbox{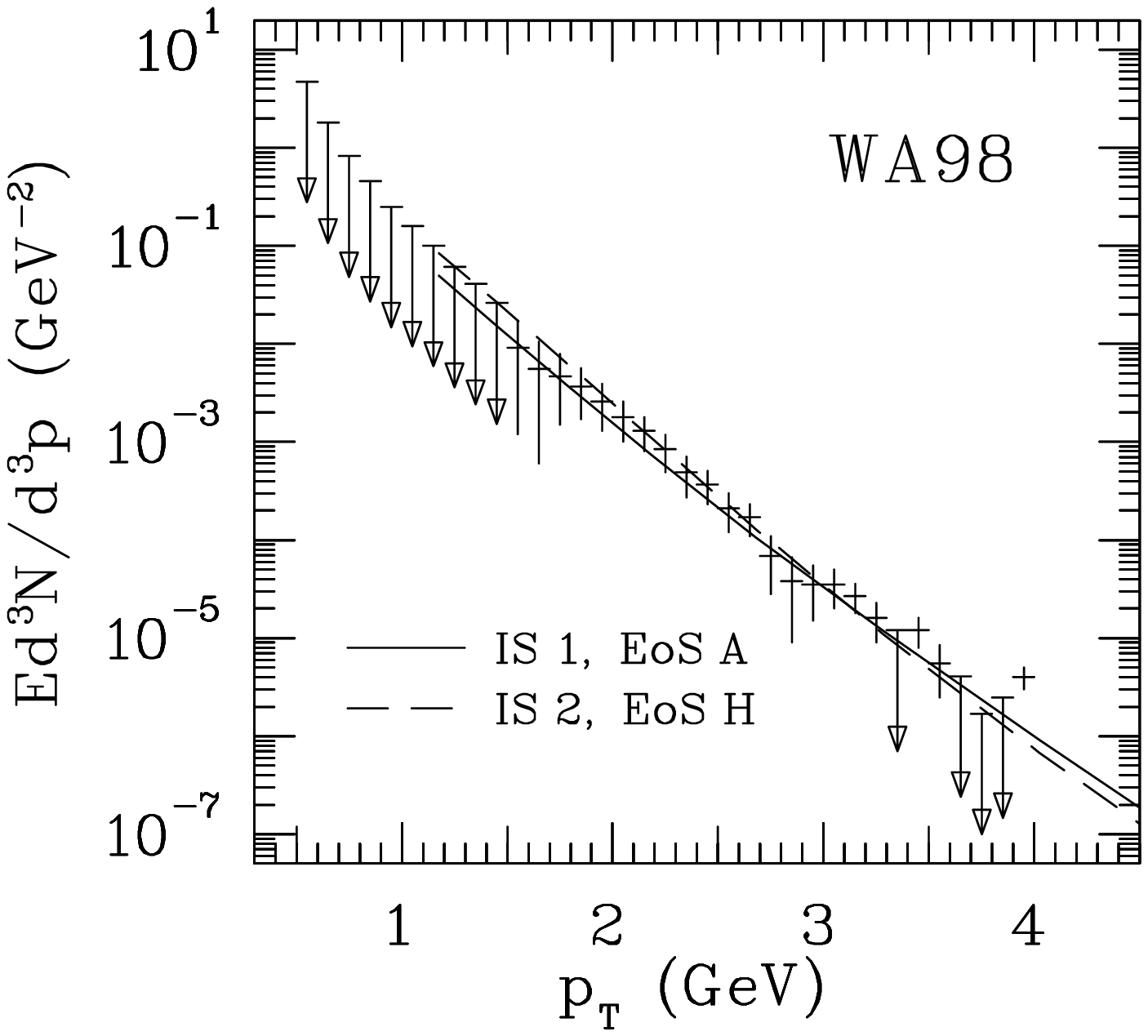}
  \caption{\protect\small
           The sum of thermal photon emission at the SPS from
           hydrodynamic calculation~\cite{sps-photons} and prompt
           photon emission~\cite{WetW} compared with WA98
           data~\cite{WA98photons}.}
   \label{fotonikuva}
 \end{minipage}
   \hfill
 \begin{minipage}[t]{63mm}
   \epsfxsize 61mm \epsfbox{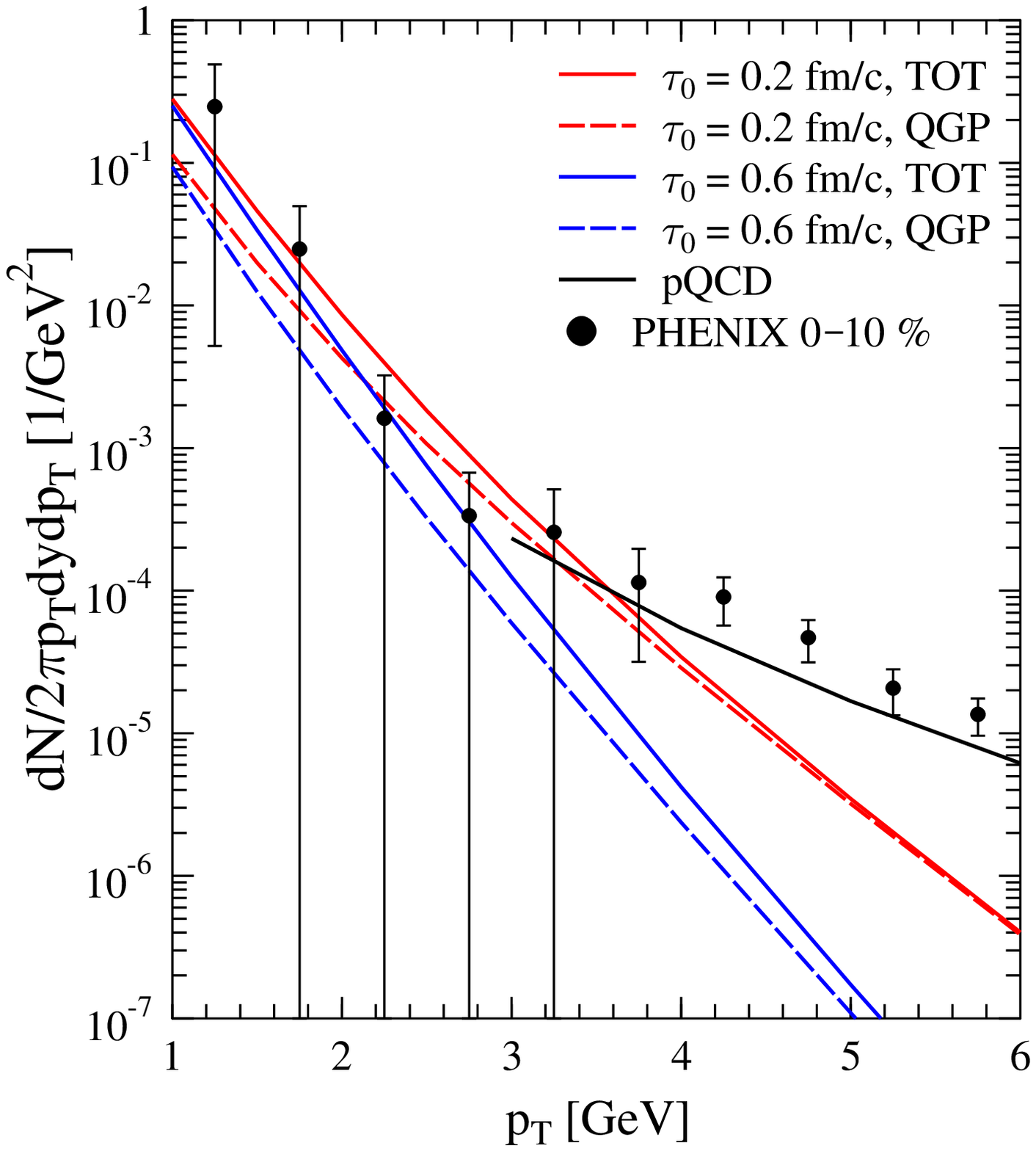}
  \caption{\protect\small
           Thermal photon emission at RHIC from hydrodynamic
           calculation using two different initial times. Solid lines
           labeled TOT indicate the total yield of thermal photons
           whereas dashed lines indicate emission from plasma. The
           pQCD calculation for prompt photons~\cite{yellow} and
           preliminary PHENIX data~\cite{Bathe:2005nz} are also
           shown.}
    \label{rhicfot}
 \end{minipage}
  \hfill
\end{figure}

To characterize the results, Figure~\ref{fotonikuva} shows the photon
spectrum calculated in Reference~\cite{sps-photons} and compares it
with WA98 data. The calculation was done using two different equations
of state (EoS A and H) and two different initial states (IS~1 and
IS~2). EoS A contains a phase transition from hadron gas to
quark-gluon plasma at $T_c=165$ MeV whereas EoS H is a purely hadronic
equation of state.  IS~1 has a very peaked initial density
distribution in the longitudinal direction, whereas IS~2 has a flatter
distribution (see Reference~\cite{Huovinen98}) and smaller maximum
temperature, which is more consistent with the assumption of hadronic
EoS. In both cases, the hadron spectra are reproduced and, as shown in
the figure, the calculated photon spectrum is within the experimental
error bars. Thus, the conclusion of the hydrodynamic studies of photon
emission at the SPS is that high temperature initial state is needed
to reproduce the measured photon spectra, but a phase transition to
plasma is not necessarily required.

\subsection{Photons at RHIC}

At the time of this writing, the situation of the photon data at RHIC
is becoming very interesting: Preliminary data
presented~\cite{Bathe:2005nz} at the latest Quark Matter 2005 meeting
indicates a clear excess of photons over decay and prompt photons in
the transverse-momentum range up to $\sim3$ GeV. Earlier measurements
have been inconclusive owing to the large error bars, but the new
method to extract the photon yield from the measurements of low-mass
$e^+e^-$ pairs appears more promising, and if the preliminary results
are confirmed by the full analysis, this data offers a long-sought
direct probe into the earliest moments of the collision.

Many authors have predicted the photon emission at RHIC and the
LHC~\cite{yellow,Sami,Gelis:2004ep,d'Enterria:2005vz,Alam,Srivastava2,
Peres2,Steffen,Hammon}. Owing to uncertainties in the initial state,
these predictions serve mostly as order-of-magnitude estimates, but
they also address the question of whether the thermal photon yield
would be larger than the prompt photon yield at any value of
$p_T$. The initial conditions of more recent studies have been
constrained to produce the total hadron multiplicity, and in
Reference~\cite{yellow}, the thermal photons are compared with the
calculated yields of decay photons both from thermal pions and prompt
pions from jet fragmentation. The measured spectrum of $\pi^0$'s is
compared with the calculations and the data is well described. The
hydrodynamical calculations with the same initial conditions are also
compared with other hadron data and the overall agreement is very
good~\cite{Eskola:2005ue}. The perturbative QCD calculation to NLO in
all quantities entering the calculation is supplemented with the
energy loss of produced jets in the thermal matter.

The main conclusions from the studies in \cite{yellow}, supported for
the hadron observables by \cite{Eskola:2005ue}, are as follows: The
understanding of the hadron spectra in terms of hydrodynamic and pQCD
calculations is quite good.  This means that the photons from hadron
decays are well under control when comparing different sources in the
calculations. The other main sources of photons are prompt photons
from primary interactions, including the photons from jet
fragmentation, and the photons from secondary interaction in produced
matter, the thermal photons\footnote{More generally the photons from
secondary interactions among the produced particles can originate also
from non-thermal processes like a high-energy quark producing photons
when Compton scattering from a lower energy thermal
gluon~\cite{Fries:2002kt}.}. These photon sources have quite distinct
transverse-momentum dependence with the cross-over from thermal
photons to prompt photons taking place at around $p_T\sim3$ GeV, the
region where the behavior of the preliminary data also changes.

The simplest hydrodynamic calculations assume a scaling expansion in
the longitudinal direction and ignore the transverse expansion. The
qfirst assumption can be argued to be reasonable in the central
rapidity region because at RHIC energy the Lorentz gamma factor is
$\sim100$, indicating a time interval on the order of 0.1 fm/$c$ for
the nuclei to pass through one another. This is shorter than the
shortest initial times used in the calculations. In the central
rapidity region, the longitudinal components are not large and the
acceleration of the longitudinal expansion is small, having little
effect on the multiplicity density or the freeze-out time at
$y\approx0$ \cite{Eskola:1997hz}. Ignoring the transverse expansion
cannot be justified, except for the photons emitted at the earliest
times from the QGP. High-$p_T$ photons from QGP are almost insensitive
to flow because they are emitted when the system is
hottest~\cite{Peres2}, but the strong flow at the late hadronic stages
enhances the emission of high-$p_T$ photons from hadron
gas~\cite{Alam}.

Predictions of the relative size of photon contributions at different
values of $p_T$ vary somewhat. For example,
Srivastava~\cite{Srivastava2} predicts that at RHIC energy, the
photons from the QGP dominate at small values of $p_T$, i.e., $p_T <
1$ GeV, and the thermal photons at high $p_T$ come mainly from the
hadronic phase.  More recent calculations lead to a conclusion that at
RHIC multiplicity the contribution from plasma dominates for $p_T >
3$, and at smaller transverse momenta the contribution from plasma and
hadron gas are the same size, with the latter slightly larger at
smallest momenta \cite{Sami,d'Enterria:2005vz}. The rates used in the
calculations is one reason for the difference: In
Reference~\cite{Srivastava2} older rates that do not include all order
of $\alpha_S$ terms are utilized, whereas the newer calculations are
based on the full order of~$\alpha_S$~\cite{Arnold:param}
results. Also, note that the dependence of the plasma contribution
depends strongly on the assumed thermalization time, $\tau_0$. When
comparing different predictions the first detail to be checked is
$\tau_0$, see Reference~\cite{yellow}.

Although the pQCD calculation of prompt photons is not entirely under
control at small transverse momenta owing to the uncertainty in the
photon fragmentation functions (see discussion in
References~\cite{yellow,Thoma}), the photon yield from secondary
collisions, i.e.\ the yield of thermal photons, decreases more steeply
and becomes negligible for $p_T\gsim4$~GeV. In the calculations
thermal and prompt photons become comparable at around $p_T\sim3$ GeV,
and because of the difference in the slopes, the uncertainty in where
the contributions cross is not large. At small momenta below 3~GeV,
thermal photons dominate \cite{Sami,d'Enterria:2005vz}, but it is not
clear whether this contribution is so large that it can be isolated
from the pion decay background.

Different contributions are compared with the preliminary photon
data~\cite{Bathe:2005nz} in Figure~\ref{rhicfot}. Here, initial times
$\tau_0 = 0.2$ and $0.6$ fm/$c$ are used in the hydrodynamical
calculation, which correspond to average initial temperatures $\langle
T \rangle = 340$ and 250 MeV, respectively. In the prompt photon
calculation no intrinsic $k_T$ is included because the same
calculation for photon production in $p+p$ collisions at RHIC
describes the data well. As indicated in Figure~\ref{rhicfot}, the
uncertainty in initial time causes more than an order of magnitude
change in the thermal photon yield at high $p_T$.

In most calculations for thermal photons at RHIC and the LHC,
chemically equilibrated matter is assumed. However, it can be expected
that the initial state is gluon dominated and quarks are suppressed
compared to their equilibrium yields~\cite{Ed}. This would lead to
smaller emission rates at a given temperature, but the suppression of
quarks means effectively smaller number of degrees of freedom and a
larger temperature for a given entropy. Detailed
calculations~\cite{Mustafa,yellow} have indicated that these two
effects largely cancel each other and that the final thermal spectra
are quite similar in both equilibrium and non-equilibrium scenarios.

\subsection{Dilepton calculations}

Like the photon measurements, the dilepton mass spectrum at the SPS
collisions measured by the CERES~\cite{ceres,ceres-S+Au} and
NA50~\cite{NA50} collaborations has been compared to hydrodynamical
calculations several times~\cite{Hung,Srivastava-lep,Sollfrank_prc,
Huovinen98,Alam,Prakash,leptons01,Kvasnikova}. The low-mass dilepton
($M_{ll} < m_{\phi}$) yield measured by CERES is dominated by emission
from the late hadronic phase (see, e.g.\ Reference~\cite{Prakash}) and
constrains only the properties of the hadronic stage of the
evolution. All these calculations agree that if meson properties in
vacuum are used~\cite{GaleLichard}, the thermal yield is not quite
sufficient to explain the observed excess. Thus, the experimental data
seems to require modifications in meson properties, but so far the
mass resolution has not been good enough to differentiate between a
change in mass and changes in spectral density. Recently, there has
been new preliminary data with better mass
resolution~\cite{Miskowiec:2005dn,Damjanovic:2005ni}, but the
conclusions are still being debated~\cite{Brown:2005ka}.

Kvasnikova et al.~\cite{Kvasnikova} addressed the intermediate-mass
($m_{\phi} < M_{ll} < m_{J/\Psi}$) dilepton yield measured by NA50. In
their calculation they found that the excess in lepton pairs in this
mass region could be explained by thermal emission in the same way as
in the low-mass region. Even if the intermediate-mass region is
expected to provide a window for observing the emission from
plasma~\cite{Shuryak:1978ij}, they found that a modest contribution
from plasma, $\sim 20$\%, was enough to fit the SPS data.

So far the genuinely hydrodynamic calculations of dilepton emission at
RHIC have been rare, and different parametrizations for the space-time
evolution of the system have been used instead~\cite{Rapp}. Medium
modifications to meson properties depend on total baryon density,
$\rho_{tot} = \rho_B + \rho_{\bar{B}}$. Because $\rho_{tot}$ at RHIC
is essentially the same as at the SPS, the low mass dilepton spectrum
at RHIC should show similar excess as seen at the SPS~\cite{Rapp}.

At intermediate masses, the thermal yield is expected to be dominated
by emission from plasma~\cite{Rapp:2004zh}. However, owing to a larger
$c\bar{c}$ production than at the SPS, the intermediate-mass dilepton
yield can be dominated by correlated charm decays, unless the
$c$-quarks rescatter significantly in the medium or can be identified
and subtracted. Isotropization of $c$-quark momentum distributions
would soften the dilepton mass spectrum, leaving a mass range in which
thermal emission dominates~\cite{Rapp:2004zh,Shuryak:1996gc}. An
additional source of lepton pairs at RHIC is the interaction of jets
with the surrounding dense plasma. According to recent
calculations~\cite{Turbide:2006mc}, the jet-plasma interactions may
dominate over thermal dilepton emission at intermediate masses.

At the time of this writing there are no calculations in which all
these contributions are taken into account and are folded with a
realistic time-evolution of the collision system. It will be
interesting to see how the future data will look and if a contribution
from plasma is needed to explain the data. The PHENIX collaboration
has recently shown the first preliminary data of low-mass dileptons at
RHIC~\cite{Toia:2005vr}, but the present experimental uncertainties
are too large to draw any conclusions. These experimental shortcomings
are currently being addressed by a detector
upgrade~\cite{Ravinovich:2005ep}.

%%%%%
%%%%%%%%%%%%%%%%%%%%%%%%%%%%%%%%%%%%%%%%%%%%%%%%%%%%%%%%%%%%%%%%%%%%%%%%
\section{CONCLUDING REMARKS}
  \label{conc}
%%%%%%%%%%%%%%%%%%%%%%%%%%%%%%%%%%%%%%%%%%%%%%%%%%%%%%%%%%%%%%%%%%%%%%%%

Hydrodynamics provides a well-defined framework to study many
experimentally accessible features of a heavy ion collision. Some
parts of the collision, such as the primary particle production of
final-state matter, lie outside hydrodynamics, and some features of
hydrodynamics, such as the freeze-out of final hadrons, have grave
uncertainties. Nevertheless, a hydrodynamic description has robust
features such as the the conservation laws, which are strictly
enforced. The main assumption of hydrodynamics, the occurrence of
frequent collisions in the final state, can also convincingly be
argued for from the large observed multiplicity. Hydrodynamics
describes the effects of collisions among the constituents, in
particular the momentum transfer between adjacent regions, in terms of
pressure that arises microscopically from momentum transfer in the
collisions. This should be a good approximation as soon as the
momentum distribution of constituents is approximately isotropic.

Hydrodynamics describes well the broadening and its mass dependence of
hadron spectra resulting from the increase of transverse collective
motion (flow). The collective motion can also be seen in the elliptic
flow. At low transverse momenta, the observed elliptic flow can be
described using hydrodynamics. Especially, the observed mass ordering
is typical for a hydrodynamic description. These are partly genuine
predictions of hydrodynamics because the amount of initial production
is fixed from the total multiplicity. Once the hadronic observables
are under control, the largest remaining uncertainty concerns the time
scales of primary production and of (approximate) thermalization.
Electromagnetic emission is very sensitive on these time scales, and
the preliminary results on photon emission at RHIC may be the first
indication that emission from the early moments of the collision can
be resolved. The emission of lepton pairs around and below the
$\phi$-meson mass, offers both a stringent test of the hydrodynamic
description of the hadron phase {and} a tool to study the effects of
medium on the properties of vector mesons. The amount of information
that can finally be obtained, depends a good deal on the progress in
experimental measurements, but it is likely that hydrodynamics will
remain an important tool for phenomenological studies for a long time.

\section*{Acknowledgments}

We would like to thank K.~Eskola, H.~Niemi, S.S.~R\"as\"anen and
T.~Hirano for discussions, H.N.~and S.S.R.~for help in preparing the
figures and T.H.~for allowing us to show some of his unpublished
results.

%%%%%%%%%%%%%%%%%%%%%%%%%%%%%%%%%%%%%%%%%%%%%%%%%%%%%%%%%%%%%%%%%%%%%%%%
%%%%%%%%%%%%%%%%%%%%%%%%%%%%%%%%%%%%%%%%%%%%%%%%%%%%%%%%%%%%%%%%%%%%%%%%

\end{document}